\documentclass[12pt]{article}
\usepackage{tocloft}
\usepackage{CJKutf8}
\usepackage{amsfonts}
\usepackage{amsmath}
\usepackage{amssymb}
\usepackage{array}
\usepackage{bigints}
\usepackage{bm}
\usepackage{booktabs}
\usepackage[nosort]{cite}
\usepackage{color}
\usepackage{dsfont}
\usepackage{float}
\usepackage{framed}
\usepackage{graphicx}
\usepackage{indentfirst}
\usepackage{mathrsfs}
\usepackage{multirow}
\usepackage{pdflscape}
\usepackage{setspace}
\usepackage{subdepth}
\usepackage{subfig}
\usepackage{titlesec}
\usepackage{wrapfig}
\usepackage[all]{xy}
\usepackage{young}
\usepackage[vcentermath]{youngtab}
\usepackage{relsize}
\usepackage{stackengine}
\usepackage{verbatim}
\usepackage{slashed}
\usepackage[utf8]{inputenc}

\usepackage{hyperref}
\hypersetup{colorlinks=true}
\hypersetup{linkcolor=black}
\hypersetup{citecolor=black}
\hypersetup{urlcolor=black}

\numberwithin{equation}{section}

\usepackage[left=2.5cm,right=2.5cm,top=2.5cm,bottom=3cm]{geometry}
\fontdimen2\font=1.2\fontdimen2{\jot}{5pt}

\titleformat{\section}{\large\bfseries}{\thesection.}{4pt}{}
\titlespacing{\section}{0pt}{20pt}{6pt}

\titleformat{\subsection}{\normalfont\bfseries}{\thesubsection.}{4pt}{}
\titlespacing{\subsection}{0pt}{15pt}{6pt}

\titleformat{\subsubsection}{\normalfont\itshape}{\thesubsubsection.}{4pt}{}
\titlespacing{\subsubsection}{0pt}{15pt}{6pt}

\titleformat{\paragraph}{\normalfont\itshape}{\theparagraph.}{4pt}{}
\titlespacing{\paragraph}{0pt}{15pt}{6pt}

\newcommand{\bea}{\begin{eqnarray}}
\newcommand{\eea}{\end{eqnarray}}
\newcommand{\beq}{\begin{equation}}
\newcommand{\eeq}{\end{equation}}

\def\<{\langle}
\def\>{\rangle}

\def\nn{\nonumber}

\def\cC {{\cal C}}

\def\cM {{\cal M}}

\def\cW {{\cal W}}

\newcommand{\mA}{\mathcal{A}}
\newcommand{\mQ}{\mathcal{Q}}

\def\cO {{\mathcal O}}

\def\cR {{\mathcal R}}

\DeclareFontShape{OT1}{cmr}{mx}{n}%
{<->cmr10}{}
\newcommand{\mytitlefont}{\fontseries{mx}\selectfont}
\DeclareMathAlphabet{\titlemath}{OT1}{cmr}{mx}{n}

\begin{document}

\begin{titlepage}

\begin{center}
			
~\\[0.7cm]
			
{\fontsize{20pt}{0pt} \mytitlefont Bootstrapping the Abelian Lattice Gauge Theories}
			
~\\[0.4cm]
\begin{CJK*}{UTF8}{gbsn}
Zhijin Li$^{\textrm{东}}$ and Shutong Zhou$^{\textrm{南}}$

~\\[0.1cm]

$^{\textrm{东}}$\,{\it Shing-Tung Yau Center and School of Physics, Southeast University , Nanjing 210096, China}~\\[0.2cm]

$^{\textrm{南}}$\,{\it Department of Physics, Nanjing University, Nanjing 210093, China}\\[0.2cm]

\end{CJK*}			
\end{center}

\vskip0.5cm
			
We study the $\mathbb{Z}_2$ and $U(1)$ Abelian lattice gauge theories using a bootstrap method, in which the loop equations and positivity conditions are employed for Wilson loops with lengths $L\leqslant L_{\textrm{max}}$ to derive two-sided bounds on the Wilson loop averages. 
We address a fundamental question that whether the constraints from loop equations and positivity are strong enough to solve lattice gauge theories.
 We answer this question by bootstrapping the 2D $U(1)$ lattice gauge theory.
 We show that with sufficiently large $L_{\textrm{max}}=60$, the two-sided bounds provide estimates for the plaquette averages with precision near $10^{-8}$ or even higher, suggesting  the bootstrap constraints are sufficient to numerically pin down this theory.
 We compute the bootstrap bounds on the plaquette averages in the 3D $\mathbb{Z}_2$ and $U(1)$ lattice gauge theories with $L_{\textrm{max}}=16$. In the regions with weak or strong coupling, the two-sided bootstrap bounds converge quickly and coincide with the perturbative results to high precision. The bootstrap bounds are well consistent with the Monte Carlo results in the nonperturbative region. 
 We observe interesting connections between the bounds generated by the bootstrap computations and the Griffiths' inequalities. We present results towards bootstrapping the string tension and glueball mass in Abelian lattice gauge theories.


\end{titlepage}

	
\setcounter{tocdepth}{3}
\renewcommand{\cfttoctitlefont}{\large\bfseries}
\renewcommand{\cftsecaftersnum}{.}
\renewcommand{\cftsubsecaftersnum}{.}
\renewcommand{\cftsubsubsecaftersnum}{.}
\renewcommand{\cftdotsep}{6}
\renewcommand\contentsname{\centerline{Contents}}
	
\tableofcontents

\newpage
\section{Introduction}
The strongly coupled gauge theories are the longstanding problems in theoretical physics. The modern conformal bootstrap program \cite{Rattazzi:2008pe, Poland:2018epd}  provides a powerful non-perturbative approach to study the conformal phases of these classical theories. 
The idea is to derive strict constraints on the parameter space of the strongly coupled gauge theories using general consistency conditions combined with the underlying positive structures. 
The bootstrap implementations are based on the correlation functions of local gauge-invariant operators.
Studies along this direction have shown that the bootstrap computations indeed can provide highly non-trivial new results for strongly coupled gauge theories \cite{Poland:2018epd}, while it is also suggested that substantially new ingredients beyond the correlators of local gauge invariant operators  are needed to improve the bootstrap results further.
  
An interesting example for strongly coupled gauge theories is provided by the 3D $U(1)$ gauge theory coupled with $N_f$ flavors of two-component Dirac fermions (QED$_3$). QED$_3$ is arguably the simplest theory which provides abundant strongly coupled physics, including the gauge confinement \cite{Polyakov:1975rs,Polyakov:1976fu,Gopfert:1981er}, chiral symmetry breaking \cite{Pisarski,Appelquist86,Dagotto:1989td} and conformal phase above the critical flavor number $N_f>N_f^*$ \cite{Appelquist88,Nash89}. QED$_3$ has played important roles in the condensed matter physics and has been proposed to have beautiful theoretical properties, such as the 3D duality web and symmetry enhancement in the infrared \cite{Senthil:2018cru}. 
An important question of QED$_3$ is to determine the critical flavor number $N_f^*$.
The conformal bootstrap results \cite{Li:2018lyb,Li:2021emd, Albayrak:2021xtd}, combined with the large $N_f$ expansion \cite{Gracey:1993sn, Borokhov:2002ib, Pufu:2013vpa, Giombi:2016fct, Chester:2016ref, Chester:2017vdh, Dupuis:2021flq} and the Monte Carlo simulations \cite{Karthik:2016ppr, Qin:2017cqw, Xu:2018wyg, Karthik:2020shl} lead to an estimate for the critical flavor number $N_f^*\in (2,4)$. Nevertheless, to bootstrap the critical indices of conformal QED$_3$, it needs extra inputs from the large 
$N_f$ perturbative expansions or the lattice simulations, while it is difficult to numerically capture the conformal gauge theories merely from bootstrap constraints. 
This observation is further supported by the fact that the four-point crossing equations in the bootstrap implementations are endowed with positive algebraic structures \cite{Li:2020bnb,Li:2020tsm}, which lead to the bootstrap bound coincidences among different global symmetries \cite{Poland:2011ey}. As a consequence, it is hard to obtain bootstrap solutions specific to the gauge theories with various global symmetries and representations. 

Non-conformal phases of the strongly coupled gauge theories have also been studied using the S-matrix bootstrap \cite{Paulos:2016fap,Kruczenski:2022lot}. The situation similar to the conformal bootstrap results has also been addressed in recent works by He and Kruczenski \cite{He:2023lyy, He:2024nwd}, that by injecting the UV information of the asymptotically free gauge theories, as well as the data on the chiral symmetry breaking in the IR, their bootstrap method for the gauge theories can provide remarkable constraints on the parameters in the intermediate energy scales in which the theories are strongly coupled. 

In summary, the conformal and S-matrix bootstrap studies suggest that the current bootstrap results on strongly coupled gauge theories depend on the extra input data from other approaches, and the bootstrap method is not full-fledged yet to provide solutions for the classical strongly coupled gauge theories. 
This situation may not be surprising, as both of the bootstrap approaches are based on the gauge invariant local operators, in which the gauge information is not directly specified and details of the gauge interactions are invisible in the bootstrap setup. 
Since the nonlocal gauge-invariant operators, e.g., the Wilson loops carry key information on the gauge interactions, 
it is natural to include the Wilson loops in the bootstrap setup. The long term goal along this direction is to develop a self-contained bootstrap method to study the strongly coupled gauge theories without resorting to additional data.  In this work we start with simpler but still physically interesting theories: the Abelian gauge theories with lattice regularization.

In lattice gauge theories, the fundamental physical observables are Wilson loops. Expectation values of the Wilson loops are restricted by the Makeenko-Migdal loop equations \cite{Makeenko:1979pb,MAKEENKO1980253,Migdal:1983qrz}. For general gauge theories the loop equations cannot be solved directly. 
In \cite{Anderson:2016rcw} Anderson and Kruczenski proposed a novel method to bootstrap the loop equations. Their key observation is that the matrices made from the inner products of the Wilson lines are positive definite. Therefore the Wilson loop averages are restricted by both the loop equations and the positive definiteness of these matrices. It turns out that the two conditions together can lead to strong constraints on the Wilson loop averages. This approach has been further improved in a recent work by Kazakov and Zheng \cite{Kazakov:2022xuh}, in which the reflection positivity \cite{Osterwalder:1977pc} in the lattice gauge theories has been employed for bootstrap computations. The authors also introduced the relaxation procedure to deal with the nonlinear parts in the loop equations.
The lattice gauge theory bootstrap is closely related to a series of bootstrap works on matrix models \cite{Lin:2020mme, Kazakov:2021lel, Koch:2021yeb, Lin:2023owt, Komatsu:2024vnb}, quantum mechanics \cite{Han:2020bkb, Berenstein:2021dyf, Bhattacharya:2021btd,Du:2021hfw, Blacker:2022szo, Berenstein:2022unr,Fan:2023tlh, Sword:2024gvv}, many body systems \cite{Nancarrow:2022wdr,Kull:2022wof} and the Ising model \cite{Cho:2022lcj}, etc.\footnote{See also the bootstrap studies of classical system \cite{Nakayama:2022ahr} and null state bootstrap without imposing positivity \cite{Li:2023nip, Li:2024rod}}

We study the  $\mathbb{Z}_2$ and $U(1)$ Abelian lattice gauge theories. The two theories can be solved analytically in 2D. The 3D $\mathbb{Z}_2$ lattice gauge theory is dual to the 3D Ising model, and provides a 3D generalization of the Kramers-Wannier duality \cite{Wegner:1971app}. It is the first known example for phase transitions without local order parameter and provides a useful model for the quantum computing \cite{Kitaev:1997wr} and generalized global symmetry \cite{Shao:2023gho}. 
The 3D $U(1)$  lattice gauge theory is one of few examples for which the confinement mechanism can be understood analytically, and it plays an instructive role in the studies of gauge confinement. 
Semi-classical analysis suggests that the 3D $U(1)$  lattice gauge theory is confined in the long-distance limit for any gauge coupling \cite{Polyakov:1975rs,Polyakov:1976fu,Gopfert:1981er}, and the confinement is triggered by condensation of the $U(1)$ monopoles. 
This theory has been extensively studied since 1970s by different approaches in order to gain detailed understanding of the confinement, see recent Monte Carlo simulations to extract the string tension and glueball spectrum \cite{Caselle:2014eka, Caselle:2016mqu, Athenodorou:2018sab}. From our point of view, the Abelian lattice gauge theories provide natural candidates to explore the bootstrap constraints from the non-local gauge invariant operators.

In this work, the first question we want to address is  whether the positivity conditions and the loop equations have sufficiently strong constraints to numerically solve the lattice gauge theories, in the sense of computing the parameters of the theories to high numerical precision. 
The previous two-sided bootstrap bounds are  rather impressive but not enough to extract precise data on the confinement, like the string tension and glueball spectrum. 
A technical obstacle is that to bootstrap the loop equations, it needs to truncate the length of the Wilson loops with a cutoff $L_{\textrm{max}}$, and the numerical results show strong dependence on the cutoff $L_{\textrm{max}}$ up to $L_{\textrm{max}}=16$
\cite{Anderson:2016rcw, Kazakov:2022xuh}.\footnote{Such dependence of the bootstrap bounds on the truncation of operators has also been shown in the bootstrap results of the Ising model on the lattice \cite{Cho:2022lcj}.}
One may expect the bootstrap results with much higher cutoff $L_{\textrm{max}}$ could  improve the results significantly. Nevertheless, the computational resource needed for the bootstrap studies increases drastically and quickly becomes impractical  with higher $L_{\textrm{max}}$.\footnote{See 
\cite{Kazakov:2024ool} for new bootstrap results with $L_{\textrm{max}}=20, 24$.} 
We find that this problem can be nicely resolved in bootstrapping 2D $U(1)$ lattice gauge theory by fixing the gauge redundancy.  The two-sided bootstrap bounds can give numerical estimates of the Wilson loop averages to numerical precision near $10^{-8}$ or even higher with $L_{\textrm{max}}=60$, thus providing strong evidence that   the loop equations and the positivity conditions contain sufficiently strong constraints to numerically solve this lattice gauge theory.

In 3D the gauge fixing does not simplify the loop equations, and we apply the bootstrap method developed in \cite{Anderson:2016rcw, Kazakov:2022xuh} for the $\mathbb{Z}_2$ and $U(1)$ lattice gauge theories. In \cite{Anderson:2016rcw, Kazakov:2022xuh} the authors focused on the $SU(N)$ lattice gauge theories in the large $N$ limit and  there are two notable differences for the Abelian lattice gauge theories studied in this work. Firstly, different from the Yang-Mills lattice gauge theories studied in \cite{Anderson:2016rcw, Kazakov:2022xuh}, the loop equations for the Abelian lattice gauge theories are linear so no need to introduce the relaxation procedure. Besides, the real sectors of the Abelian lattice gauge theories satisfy Griffiths' inequalities \cite{GriffithsI,Griffiths69,Ginibre1970}. The bootstrap study of the Griffiths' inequalities was firstly introduced in \cite{Cho:2022lcj}. The Griffiths' inequalities can be used to prove the convergence of the lattice system in the infinite volume limit. They have been employed to derive various interesting bounds on the lattice theories nearly half a century before the modern bootstrap endeavors.
We will show that they have interesting connections with our bootstrap bounds.

This paper is organized as follows. In Section \ref{sec2} we show the loop equations of the $U(1)$ lattice gauge theories. For the $\mathbb{Z}_2$ lattice gauge theory, the loop equations are replaced by the link-flip equations. 
In Section \ref{sec3} we explain the positivity conditions, including the square positivity and the reflection positivity. In addition, we discuss Griffiths' inequalities in Abelian lattice gauge theories and its subtlety in the $U(1)$ lattice gauge theory. 
In Section \ref{sec4} we bootstrap the 2D $U(1)$ lattice gauge theory and compare the results with the analytical solution. We particularly focus on the convergence of the two-sided bounds with increasing cutoff $L_{\textrm{max}}$. 
In Section \ref{sec5} we bootstrap the 3D  $U(1)$ and $\mathbb{Z}_2$ lattice gauge theories, and compare the two-sided bootstrap bounds with the Monte Carlo results and the perturbative expansions in the strong and weak coupling limits. We discuss the bootstrap estimations of the string tension and glueball mass.
We summarize the results in Section \ref{sec6} and discuss the future bootstrap studies of the lattice gauge theories.

\vspace{3mm}

\noindent {\bf Note added:} During the course of this work we learned of the ongoing paper \cite{Kazakov:2024ool} which bootstraps the $SU(N)$ lattice Yang-Mills gauge theories with finite $N$s. The authors also apply gauge fixing to simplify the 2D loop equations and obtain two-sided bounds with high precision for non-Abelian lattice gauge theories, reminiscent to our results in Section \ref{sec4}.
We thank the authors for coordinating
submission.

\section{Loop equations in Abelian lattice gauge theories} \label{sec2}

 The expectation values of the Wilson loops are constrained by the Schwinger-Dyson equations which can be obtained from the Wilson action of the lattice gauge theory by taking infinitesimal shift of the link variables of the lattice \cite{Makeenko:1979pb,MAKEENKO1980253, Migdal:1983qrz}. While for the lattice gauge theories with discrete gauge group, e.g., $\mathbb{Z}_2$, the link variables can only take discrete values which do not admit infinitesimal shift. In this case, the loop equations are instead obtained by taking permutations of the link variables, or flipping the sign of the link variables in the $\mathbb{Z}_2$ lattice gauge theory.

\subsection{Loop equations in  $U(1)$ lattice gauge theories}
We study the $U(1)$ gauge theories in an infinite hypercubic lattice in a general dimension $D$.\footnote{In 2D, the loop equations can be simplified after gauge fixing, see Section \ref{sec4}} In lattice gauge theories, an oriented link  from site $x$  to $x+\mu$  corresponds to a dynamical variable $U_\mu(x)$ valued in the gauge group $G$: $U_\mu(x)\in G$. For the $U(1)$ lattice gauge theory, it is a phase $U_\mu(x)=e^{i\,\theta_\mu(x)}$, and its complex conjugate denotes the same link with reversed direction: $U^*_\mu(x+\mu)=e^{-i\,\theta_\mu(x)}$. The action of this theory is given by the Wilson action \cite{Wilson:1974sk}
\beq
S=-\frac{1}{\lambda}\sum\limits_{P} U_P\rightarrow -\frac{2}{\lambda}\sum\limits_{P} \cos{\theta_P}, \label{U1action}
\eeq
where $P$ denotes the oriented plaquettes consisting of four links, and the sum is over all plaquettes of the lattice. The variable $U_P$ represents the product of the link variables around the plaquette $P$: 
$$U_P\equiv U_\mu(x)U_\nu(x+\mu)U_\mu^*(x+\nu)U_\nu^*(x),$$
and for the $U(1)$ lattice gauge theory, it is $$ U_P=e^{i\,\theta_P},~~~~ \theta_P\equiv \theta_{\mu\nu}(x)=\theta_\mu(x)+\theta_\nu(x+\mu)-\theta_\mu(x+\nu)-\theta_\nu(x).$$ The parameter $\lambda$ is dimensionless and relates to the gauge coupling $e$ and lattice spacing $a$ through  $\lambda=a e^2$.

The partition function of the $U(1)$ lattice gauge theory is
\beq
Z_\lambda=\int \prod\limits_{\{x,\mu\}} dU_\mu(x)\,e^{-S}=\int_{0}^{2\pi}\prod\limits_{\{x,\mu\}}d\theta_{\mu}(x)e^{\frac{2}{\lambda}\sum\limits_{P} \cos{\theta_P}}. \label{U1Partition}
\eeq
The physical observable is the Wilson loop $\cW_\cC$ along a closed path $\cC$:
$\cW_\cC\equiv e^{i\theta_\cC}, ~\theta_\cC=\sum\limits_{\ell\in\cC}\theta_\ell$. Its expectation value is given by
\beq
\langle \cW_\cC\rangle=\frac{1}{Z_\lambda}\int \prod\limits_{\{x,\mu\}} dU_\mu(x) \, \cW_\cC \,e^{-S}=\int_{0}^{2\pi}\prod\limits_{\{x,\mu\}}d\theta_{\mu}(x) \,  \exp{\left(\frac{2}{\lambda}\sum\limits_{P} \cos{\theta_P}+   i\,\theta_\cC \right)}. \label{WLaverage}
\eeq
Above integral can be evaluated using $1/\lambda$ expansion from which the confined phase in the strongly coupled limit can be analytically confirmed \cite{Wilson:1974sk}. 
A more physically relevant problem is the confinement in the continuum limit, and it needs to show the expectation value of the large Wilson loops in the limit $\lambda\rightarrow 0$, for which a nonperturbative method is needed. 
Traditionally, this has been resorted to Monte Carlo simulations. The bootstrap method suggests that instead of solving the integrals over the link variables $U_\mu(x)$, it may suffice to focus on the gauge invariant operators only and find solutions of the Wilson loop averages $\langle \cW_\cC\rangle$ from the equations and positivity conditions they should satisfy. 

The loop equations of $\langle \cW_\cC\rangle$ can be derived through the standard procedure for the Schwinger-Dyson equations \cite{Makeenko:1979pb, MAKEENKO1980253,Migdal:1983qrz}. 
Consider the Wilson loop along a closed path $\cC$ with links $\{\mu\nu\rho\dots\}\equiv\{\mu\hat{C}\}$, in which the starting and ending point is implicitly assumed at $x$, while $\mu, \nu,\rho \in \{1,2,\dots,D\}$ indicate the directions of the links and $\Bar{\mu}=-\mu$ denotes the opposite link direction $\mu$. Since we study the Abelian gauge theories, the order of the links does not matter. 
Note the path $\cC$ can be redundant if the path passes the same link twice with opposite directions, e.g., 
\beq
\cC'\equiv\{\mu\nu\Bar{\nu}\Bar{\mu}\hat{C}\}=\{\hat{C}\}\equiv \cC. \label{linkcancel}
\eeq
The first path $\cC'$ gives the same Wilson loop but the variation of its first link $U_\mu(x)$ can generate new loop equations, as will be discussed soon. We call a loop path $\cC$ is {\it reduced} if all the redundancies are removed. 
Now denote the average value of a Wilson loop by
\beq
W[\mu\hat{C}]=W[\cC]\equiv\langle \cW_\cC\rangle. 
\eeq
The path integral (\ref{WLaverage}) is invariant under the variation of the link variable $U_\mu\rightarrow (1+i\epsilon )U_\mu$, which leads to the constraints on the averages of several related Wilson loops
\beq
\sum\limits_{|\alpha|\neq \mu}W[\mu\alpha\Bar{\mu}\Bar{\alpha}\mu\hat{C}] -W[\alpha{\mu}\Bar{\alpha}\hat{C}] + n \lambda W[\mu\hat{C}]=0,     \label{lpequation}  
\eeq
where the variable $\alpha$ takes all the $2(D-1)$ directions except $\pm \mu$. The parameter $n$ is the repetition number of the link $U_\mu(x=0)$ in $\cC$. Above equations are linear for any Wilson loops, which is the main difference between the loop equations of the $U(1)$ and $SU(\infty)$ lattice gauge theories \cite{Anderson:2016rcw, Kazakov:2022xuh}. For the latter case, the loop equations are non-linear and cannot be directly studied using linear programming unless one introduces the relaxation procedure \cite{Kazakov:2021lel, Kazakov:2022xuh}. 

Depending on the geometrical relation between the shifted link $U_\mu(x)$ and the {\it reduced} Wilson loop $\cC$, one can generate three different types of loop equations:

\begin{enumerate}
    \item  If the shifted link belongs to the {\it reduced} Wilson loop, then $n>0$ and the loop equation contains the gauge coupling $\lambda$. The majority of the Wilson loops appear in the equation are connected Wilson loops.

    \item Shifted link does not belong to but connect to the {\it reduced} Wilson loop, then $n=0$ and the loop equation is independent of the gauge coupling $\lambda$. The majority of the Wilson loops appear in the equation are connected Wilson loops. These loop equations are dubbed ``backtrack loop equations".

    \item Shifted link is away from the {\it reduced} Wilson loop, then $n=0$ and the loop equation is independent of the gauge coupling $\lambda$. Moreover, the Wilson loops appear in the equation are disconnected Wilson loops. These loop equations are new for the $U(N)$ lattice gauge theories with finite $N$, in which the disconnected Wilson loops are not factorized and can provide key information to study glueball spectrum, etc.
    
\end{enumerate}


\subsection{Link-flip equations in  $\mathbb{Z}_2$ lattice gauge theories}
Now we consider the equations of the Wilson loop averages in the $\mathbb{Z}_2$ lattice gauge theory. Denote the link variable by $\sigma_\mu(x)$. The theory is real so the link is non-oriented:, $\sigma_\mu(x)=\sigma^*_\mu(x)=\sigma_{\bar{\mu}}(x+\mu)$. Since the link variable only has two choices $\sigma_\mu(x)=\pm1$, it does not admit an infinitesimal shift and one has to consider finite variation of $\sigma_\mu(x)$ to derive the loop equation. The method is reminiscent to the spin-flip equation for the Ising model without external magnetic field, which has been studied in detail in \cite{Cho:2022lcj}.

The action of the $\mathbb{Z}_2$ lattice gauge theory is
\beq
S=-J\sum\limits_P \sigma_P\equiv -J\sum\limits_P \sigma_\mu(x)\sigma_\nu(x+\mu)\sigma_\mu(x+\nu)\sigma_\nu(x). \label{2DZ2action}
\eeq
The expectation value of a Wilson loop $\cW_\cC=\prod\limits_{\ell\in\cC} \sigma_\ell\equiv \sigma_\cC$ along a path $\cC$ 
is given by
\beq
W[\cC]\equiv\langle \cW_\cC\rangle =\frac{1}{Z_J}\sum\limits_{\sigma_i} \sigma_\cC \, \exp{\left(J\sum\limits_P \sigma_P\right)},
\eeq
where $Z_J$ is the partition function of the $\mathbb{Z}_2$ lattice gauge theory.
The expectation value $W[\cC]$ is invariant by flipping the sign of the link variable $\sigma_\mu(x)\rightarrow -\sigma_\mu(x)$, leading to the equation
\beq
(1-2\chi_\cC(\sigma_\mu(x)))W[\cC]=\frac{1}{Z_J}\left\langle \sigma_\cC \exp{\left(-2J\sigma_\mu(x)\sum\limits_{|\nu|\neq\mu} \sigma_\nu(x+\mu)\sigma_\mu(x+\nu)\sigma_\nu(x)\right)} \right\rangle, \label{flipeq}
\eeq
where $\chi_\cC(\sigma)$ depends on the position of the flipped link variable $\sigma_\ell$
\beq
\chi_\cC(\sigma_\ell)=  \left\{ \begin{array}{rcl}
         1, & ~~\ell\in\cC,
          \\ 0,  & ~~\ell\notin \cC. 
                \end{array}\right.
\eeq
The exponential function in (\ref{flipeq}) can be reduced to a polynomial function due to the discreteness of the link variable $\sigma$. Specifically the composite variable
\beq
w\equiv \sum\limits_{|\nu|\neq\mu}\sigma_\nu(x+\mu)\sigma_\mu(x+\nu)\sigma_\nu(x) \in \{0,\pm2,\cdots,\pm 2(D-1)\}
\eeq
satisfies the following algebraic constraint
\begin{equation}
    \prod\limits_{k=-2(D-1)}^{2(D-1)}(w-2k)=0.
\end{equation}
In consequence, the higher order functions $w^n,~n\leqslant2D-1$ can be expanded in terms of $w^\ell$ with $\ell<2D-1$
\begin{equation}
    w^n=\sum\limits_{\ell=0}^{\ell=2(D-1)}p_{n,\ell}w^\ell. \label{wn}
\end{equation}
In 2D, the coefficients $p_{n,\ell}$ are
\begin{equation}
    p_{2k,2}=4^{k-1},~~~~p_{2k-1,1}=4^{k-1}, ~~~~p_{2k-1,2}=p_{2k,1}=0.
\end{equation}
In 3D, the coefficients $p_{n,\ell}$ are
\begin{align}
    p_{2k,2}&=\frac{1}{3}\left(2^{2k}-2^{4k-4}\right),~~~~~\;\;p_{2k,4}=\frac{1}{3}\left(2^{4k-6}-2^{2k-4}\right), \nn\\
    p_{2k+1,1}&=\frac{1}{3}\left(2^{2k+2}-2^{4k}\right),~~~~p_{2k+1,3}=\frac{1}{3}\left(2^{4k-2}-2^{2k-2}\right), \nn
\end{align}
while all the other coefficients vanish.
Now one can expand the exponential function in (\ref{flipeq}) and introduce the $w^n$ reduction formula (\ref{wn}), through which the equation (\ref{flipeq}) becomes
\begin{align}
    2\chi_\cC \langle\sigma_\cC \rangle+\sum\limits_{\ell=0}^{2D-2}f_\ell \langle \sigma_\cC w^\ell\rangle -
    \sum\limits_{\ell=0}^{2D-2}g_\ell \langle \sigma_\cC w^\ell\sigma_\mu(x) \rangle=0. \label{z2loopeq}
\end{align}
The coefficients $f_\ell, ~g_\ell$ depend on the lattice dimension $D$. In 2D, the coefficients are
\begin{equation}
    f_1=g_2=0,~~~f_2=\frac{1}{4}(\cosh{4J}-1), ~~g_1=\frac{1}{2}\sinh{4J},
\end{equation}
while in 3D, the non-vanishing coefficients are 
\begin{align}
    f_2&=\frac{1}{48}\left(-15+16\cosh{4J}-\cosh{8J} \right),~~ f_4=\frac{1}{192}\left(3-4\cosh{4J}+\cosh{8J} \right), \nn\\
    g_1&=\frac{1}{12}\left(8\sinh{4J}-\sinh{8J} \right),~~~~~~~~~~~\,    g_3=\frac{1}{48}\left(-2\sinh{4J}+\sinh{8J} \right). \nn
\end{align}

Above loop equations share several similarities with the spin-flip equations of the Ising model on a lattice \cite{Cho:2022lcj}. In 3D, they lead to essentially the same solution, as the 3D $\mathbb{Z}_2$ lattice gauge theory is dual to the 3D Ising model. In 2D they have completely different solutions. For the $\mathbb{Z}_2$ lattice gauge theory, by taking the temporal gauge, i.e., all the links in the temporal direction $\mu=\tau$ are set to unit: $\sigma_{\tau}(x)=1$, the Wilson action (\ref{2DZ2action}) becomes the action of
the 1D Ising model, and the Wilson loops with $n$ plaquettes have the factorized form $w_1^n$, where $w_1$ is the plaquette average. The link-flip equation (\ref{z2loopeq}) of the plaquette is
\begin{equation}
    -\frac{1}{2}\sinh{(4J)}+(1+\cosh{(4J)})w_1-\frac{1}{2}\sinh{(4J)}w_1^2=0. \label{Z2loopeq}
\end{equation}
which is simply solved by
   $ w_1=\tanh{J}$, the solution of the 1D Ising model.


\section{Positivity conditions and Griffiths' inequalities} \label{sec3}
In this section we explain the positivity properties of the loop equations.
Generally, the lattice gauge theory contains infinite loop equations with an infinite set of Wilson loop averages, so they are quite difficult to solve directly. The exceptional examples are the loop equations in 2D, e.g., the $SU(\infty)$ lattice gauge theory \cite{Gross:1980he, Wadia:2012fr}. The bootstrap method aims to extract constraints on the solutions from positivity conditions associated with the loop equations.  

\subsection{Square positivity and reflection positivity}
A critical positivity condition, dubbed square positivity in this work,\footnote{In \cite{Anderson:2016rcw, Kazakov:2022xuh}, the square positivity has a different name. Here we follow the terminology in \cite{Cho:2022lcj} which studied the Ising model on the lattice.} is from Anderson and Kruczenski \cite{Anderson:2016rcw}. They considered the vector space spanned by the Wilson line operators $\cW_i$ along the paths $\cC_i$, which share the same starting and ending points ($x_1\rightarrow x_2$). An arbitrary operator in the vector space is given by
\begin{equation}
    \cW=\sum\limits_i c_i\cW_i.
\end{equation}
Then its inner product is non-negative for any $c_i$:
\begin{equation}
    \langle \cW|\cW\rangle=\sum_{i,j}c_i^* \,\cM_{ij} \, c_j\geqslant0, ~~~~\cM_{ij}\equiv\langle \cW_i^*|\cW_j \rangle, \label{sqmatrix}
\end{equation}
and the matrix $\cM$ is positive semi-definite. The conjugate action on the Wilson line $\cW_i^*$ switches its starting and ending points $x_1\leftrightarrow x_2$ and reverses the orientation of each link. Its inner product with the Wilson line $\cW_j$ gives a Wilson loop around the path $x_2\rightarrow \cC^*_i \rightarrow x_1 \rightarrow \cC_j\rightarrow x_2$.  
It has been shown that, by considering the vector space spanned by only two or three Wilson lines, the positive semi-definiteness condition of their product matrix $\cM$ can lead to interesting constraints \cite{Anderson:2016rcw}. The constraints can be significantly improved by introducing more Wilson lines in the vector space \cite{Anderson:2016rcw, Kazakov:2022xuh}.

Another positive constraint is reflection positivity \cite{Osterwalder:1977pc}, which has been implemented to bootstrap lattice gauge theories in \cite{Kazakov:2022xuh} and Ising model in \cite{Cho:2022lcj}. The reflection positivity corresponds to the unitarity of the lattice field theories. 
The reflection operation $\Theta$ of an operator is defined as follows. Firstly, one chooses a direction $\rho$ as the ``time" direction. The Hilbert space $\mathcal{H}$ is separated into two pieces with $\rho>\rho_0$ ($\mathcal{H}^+$) or $\rho<\rho_0$ ($\mathcal{H}^-$). Then the reflection operation  $\Theta: \mathcal{H}^+\rightarrow \mathcal{H}^-$ is an anti-linear map 
\begin{equation}
   \Theta \cdot\cW_\cC= \cW^*_{R\cdot \cC},
\end{equation}
where $R\cdot \cC$ is the reflection of the oriented path $\cC$ with respect to the plane $\rho=\rho_0$. For the cubic lattice, there are three possible reflection planes that can map $\mathcal{H}^+$ to $\mathcal{H}^-$: the ``time" direction $\rho$ is parallel to one side of the lattice, and the reflection plane is at $\rho=0$ or $\rho=\frac{1}{2}$; or alternatively, the ``time" direction $\rho$ is along a diagonal direction of the lattice. See \cite{Kazakov:2022xuh, Cho:2022lcj} for more details.

\subsection{Griffiths' inequalities in Abelian lattice gauge theories} \label{sec3p2}
The Griffiths' inequalities were first proved for the ferromagnetic Ising model \cite{GriffithsI}. A fundamental application of the Griffiths' inequalities is the proof of the convergence of the Ising model correlation functions in the infinite volume limit \cite{GriffithsI}. Their roles in bootstrap have been studied in \cite{Cho:2022lcj}. Here we focus on the Griffiths' inequalities in the Abelian lattice gauge theories. 

Besides the Ising model, the Griffiths' inequalities have been known to hold in many lattice models \cite{Kelly68, Griffiths69, Ginibre1970}. Especially in \cite{Ginibre1970}, Ginibre has proved the Griffiths' inequalities in sufficient generality. Here we explain Ginibre's proof related to the Abelian lattice gauge theories. We clarify a subtlety due to which the inequalities are not always satisfied, and for some cases, even the direction of the inequality can be switched. 

Consider a compact space $X$. The algebra of complex continuous functions on $X$ is $\mA$. The elements in $X$ are denoted by $x,~y, \dots$ and $\mu$ is a positive measure of total mass $1$ on $X$. We are interested in a particular convex subset $\mQ\subset\mA$ that satisfies the following.

\begin{enumerate}
    \item For any $p(x)\in\mQ,~q(x)\in \mQ$ and constants $a\geqslant0,~b\geqslant0$, the functions
    \begin{equation}
        ap(x)+bq(x), p(x)q(x),p^*(x)\in\mQ. \label{Q1}
    \end{equation}
    \item For any finite set of functions $p_i(x)\in\mQ$, the following condition is always fulfilled
\begin{equation}
    \int d\mu(x)d\mu(y)\prod\limits_{i=1}^n(p_i(x)\pm p_i(y))\geqslant0. \label{Q3}
\end{equation}
\end{enumerate}
The convex cone $\mQ$ has following interesting properties. 

\vspace{3mm}

{\bf Proposition 1. } 
For any finite set of functions $p_i(x)\in\mQ$, the integral is non-negative
\begin{equation}
   \int d\mu(x) \prod\limits_i p_i(x)\geqslant0. \label{prop1}
\end{equation}

{\it Proof.}  Proposition 1 is a direct consequence of the presumed condition (\ref{Q3}). 
One firstly rewrites the above integral 
\begin{equation}
    \int d\mu(x) \prod\limits_{i=1}^n p_i(x)=\frac{1}{2}\int d\mu(x) d\mu(y)\left( \prod\limits_{i=1}^n p_i(x)+\prod\limits_{i=1}^n p_i(y)\right),
\end{equation}
then the integrand can be  expanded iteratively through the relation
\begin{align}
    \prod\limits_{i=1}^n p_i(x)+\prod\limits_{i=1}^n p_i(y)&= \nn\\
   &\hspace{-2.5cm} \frac{1}{2}[p_1(x) +p_1(y)]\left( \prod\limits_{i=2}^n p_i(x)+\prod\limits_{i=2}^n p_i(y)\right)+\frac{1}{2}[p_1(x)-p_1(y)]\left( \prod\limits_{i=2}^n p_i(x)-\prod\limits_{i=2}^n p_i(y)\right) \nn\\
    &=\dots=\sum\limits_{k} c_k \left[\prod\limits_{i=1}^n(p_i(x) \pm p_i(y))\right] \label{proddecomp}
\end{align}
with positive coefficients $c_k>0$. Due to (\ref{Q3}), the integral of each term in the expansion (\ref{proddecomp}) is positive, which proves Proposition 1.

Now assume $X$ is the phase space of a physical system described by an action $S(x)$: $-S(x)\in\mQ$, and the partition function $Z_S$ is non-vanishing. 

\vspace{3mm}

{\bf Griffiths' first inequality.} If a physical observable $p(x)\in\mA$ belongs to the convex cone $p(x)\in\mQ$, then its expectation value $\langle p(x)\rangle\geqslant0$.

{\it Proof.} Since $-S(x), p(x)\in\mQ$, according to the definition of the convex cone, we have $e^{-S(x)}\in \mQ$ and $p(x)\,e^{-S(x)}\in\mQ$. Then the Proposition 1 guarantees that the two integrals are non-negative
\begin{equation}
    Z_S=\int d\mu(x)\, e^{-S(x)}>0, ~~~ \langle p(x)\rangle =\frac{1}{Z_S}\int d\mu(x) \,p(x) \,e^{-S(x)}\geqslant0. 
\end{equation}

{\bf Griffiths' second inequality.} 
For any $p(x), q(x)\in\mQ$, the expectation values satisfy $\langle pq\rangle-\langle p\rangle \langle q\rangle\geqslant0$.

{\it Proof.} The  difference between the expectation values can be expanded as
\begin{align}
    \langle pq\rangle-\langle p\rangle \langle q\rangle =\frac{1}{2Z_S(x)^2}\int d\mu(x)d\mu(y)\left(p(x)-p(y)\right)\left(q(x)-q(y)\right)e^{-S(x)-S(y)} \nn\\
    =\frac{1}{2Z_S(x)^2} \sum\limits_{i=0}^\infty \frac{1}{i!}\int d\mu(x)d\mu(y)\left(p(x)-p(y)\right)\left(q(x)-q(y)\right)\left(-S(x)-S(y)\right)^i. \label{Griffiths2}
\end{align}
Here one assumes that the functions $p,q, S$ have sufficient good properties so that the order of the series expansion and the integration can be switched.\footnote{For finite size lattice models, this is obviously true. } The Griffiths' second inequality follows from the fact that each term in (\ref{Griffiths2}) is positive.

Now we study the Griffiths' inequalities for the Abelian lattice gauge theories. We start with the $U(1)$ theory, for which one needs to know the convex cone $\mQ$ spanned by the Wilson action (\ref{U1action}) $S\propto -\sum\limits_P \cos{\theta_P} $ and Wilson loops $\cW_\cC=e^{i\theta_\cC}$. 
It turns out that not all but the real part of the Wilson loops, $\cos{\theta_\cC}$, are in the convex cone $\mQ$. 

\vspace{4mm}

{\bf Proposition 2.}
The convex cone $\mQ_{U(1)}$ generated by functions of the link variables $\theta_\ell, \ell\in\mathbb{Z}^D$
\begin{equation}
    \mQ_{U(1)}=\{\cos{ n_\ell \theta_\ell}|,n_\ell\in\mathbb{Z},\ell\in\mathbb{Z}^D\}   \label{convexU1}
\end{equation}
satisfies the conditions (\ref{Q1}, \ref{Q3}).

{\it Proof.} The condition (\ref{Q1}) can be proved by a simple fact:
\begin{equation}
2\cos{(n_\ell\theta_\ell)}\cdot\cos{(n'_{\ell}\theta_{\ell})}= \cos{((n_\ell+n'_{\ell})\theta_\ell)}+\cos{((n_\ell-n'_{\ell})\theta_\ell)}\in \mQ_{U(1)}. \label{prod2sum}
\end{equation}
So the convex cone $\mQ_{U(1)}$ is multiplicative and satisfies (\ref{Q1}).

The condition (\ref{Q3}) requires that for any finite set of functions in $\mQ_{U(1)}$: $p_i(\theta)=\sum\limits_i a_{i} \cos{n_{i,\ell} \theta_\ell},~i=1,2,\dots,n$, the following integral is non-negative
\begin{equation}
    \int d\theta_\ell d\theta'_\ell\prod\limits_{i=1}^n\left(     p_i(\theta)\pm p_i(\theta')\right)\geqslant0. \label{Q3U1}
\end{equation}
Now using the relations
\begin{equation}
    \label{cos1}
\begin{split}
    \cos{n_\ell\theta_\ell}+\cos{n_\ell\theta'_\ell}&=2\cos{n_\ell\frac{\theta_\ell-\theta'_\ell}{2} }\cos{n_\ell\frac{\theta_\ell+\theta'_\ell}{2}}, \\
    \cos{n_\ell\theta_\ell}-\cos{n_\ell\theta'_\ell}&=2\sin{n_\ell\frac{\theta_\ell-\theta'_\ell}{2} }\sin{n_\ell\frac{\theta_\ell+\theta'_\ell}{2}},  
\end{split}
\end{equation}
the integrand in (\ref{Q3U1}) can be decomposed into the products of the functions in the RHS of (\ref{cos1}) with positive coefficients, and each term leads to the integral
\begin{equation}
    \int d\theta_\ell d\theta'_\ell \;P\left(\frac{\theta_\ell+\theta'_\ell}{2}\right)P\left(\frac{\theta_\ell-\theta'_\ell}{2}\right), \label{P2s}
\end{equation}
where the functions $P(\bar{\theta}_\ell)$ consist of the factors $\cos{n_\ell\bar{\theta}_\ell}/\sin{n_\ell\bar{\theta}_\ell}$ so are periodic in $\bar{\theta}_\ell$ with period $2\pi$. Take the variable transformations:
\begin{equation}
    \cR: \phi_\ell=\frac{\theta_\ell+\theta'_\ell}{2}, ~~~~~\phi'_\ell=\frac{\theta_\ell-\theta'_\ell}{2},
\end{equation}
which maps the square region $\theta_\ell\in [-2\pi,2\pi), \theta'_\ell\in [-2\pi,2\pi)$ into, after taking the periodicity of $\phi_\ell/\phi'_\ell$ into account, the double cover of a fundamental domain $\phi_\ell\in[-\pi, \pi),\phi'_\ell\in[-\pi, \pi)$, thus the integral (\ref{P2s}) can be written into the form
\begin{equation}
    \int d\phi_\ell \,d\phi'_\ell \;P(\phi_\ell) P(\phi'_\ell) = \left(\int d\phi_\ell\;P(\phi_\ell)\right)^2\geqslant0,
\end{equation}
which confirms that the integral (\ref{Q3U1}) is non-negative.  This completes the proof of Proposition 2.
    
\vspace{3mm}

{\bf Remarks.} It is important to note that the convex cone $\mQ_{U(1)}$ is expanded by the functions $\cos{n_\ell \theta_\ell}$ while the general Wilson loops $\cW_\cC=\exp{\left(i\theta_\cC\right)}$ in  $U(1)$ lattice gauge theory contain extra imaginary parts so $\cW_\cC\notin \mQ_{U(1)}$. Nevertheless, the expectation values
of the Wilson loops are always real. The imaginary part is odd under $\theta_\ell\rightarrow -\theta_\ell$ while the Wilson action is even, so the imaginary part vanishes in the average
\begin{equation}
   \langle \cW_\cC\rangle= \left\langle \cos{\theta_\cC}\right\rangle\geqslant0, \label{sgWL}
\end{equation}
therefore the expectation values of the Wilson loops inherit the properties of the convex cone $\mQ_{U(1)}$, e.g., non-negative. However, there are exceptions related to the product of Wilson loops. Consider a new Wilson loop from product of the two $\cW_{\cC}$: $\cW_{\cC^2}=\exp{\left( 2i\theta_\cC\right)}$, and its expectation value is
\begin{equation}
    \langle \cW_{\cC}\cW_{\cC}\rangle=\langle \cW_{\cC^2}\rangle= \left\langle \cos{2\theta_\cC}\right\rangle=\left\langle \cos^2{\theta_\cC}\right\rangle- \left\langle\sin^2{\theta_\cC}\right\rangle. \label{wcwc}
\end{equation}
The Griffiths' second inequality suggests
\begin{equation}
    \left\langle \cos^2{\theta_\cC}\right\rangle\geqslant \left\langle \cos{\theta_\cC}\right\rangle^2. 
\end{equation}
Note above inequality does not imply the possible Griffiths' second inequality for the Wilson loops
\begin{equation}
    \langle \cW_{\cC}\cW_{\cC}\rangle \ngeqslant\langle \cW_{\cC}\rangle^2,  
\end{equation}
or more generally
\begin{equation}
    \langle \cW_{\cC_1}\cW_{\cC_2}\rangle \ngeqslant\langle \cW_{\cC_1}\rangle\langle \cW_{\cC_2}\rangle.  
\end{equation}
The reason is that the imaginary part of $\cW_\cC$ also contributes to the product of Wilson loops,  which makes the Wilson loops $\cW_\cC$ run outside of the convex cone $\mQ_{U(1)}$, thus the Griffiths' second inequalities are not sustained. In fact, in Section \ref{sec4p2} we will show that the direction of the inequality can be reversed in 2D $U(1)$ lattice gauge theory!

In contrast, for the $\mathbb{Z}_2$ lattice gauge theory, all the Wilson loops are real and belong to the convex cone $\mQ_{\mathbb{Z}_2}$ which is a $\mathbb{Z}_2$ truncation of $\mQ_{U(1)}$. Therefore, the two Griffiths' inequalities are satisfied.
 
\subsection{A universal lower bound from Griffiths' inequalities}
The convex cone $\mQ_{U(1)}$, though does not cover the Wilson loop space, has interesting applications \cite{DeAngelis:1977sr}: it can prove the existence of the infinite volume limit of the in $U(1)$ lattice gauge theories, and moreover, it provides a lower bound on the averages of the Wilson loops $\langle \cW^{D}_\cC\rangle_D$ in a general dimension $D>2$:\footnote{This bound will be referred to as Griffiths lower bound, as it is a direct consequence of the Griffiths' inequalities, though the bound was firstly proposed in \cite{DeAngelis:1977sr}.}
\begin{equation}
  \langle \cW_\cC\rangle_D\geqslant \langle \cW_\cC\rangle_{D=2}.\label{Gribd}
\end{equation}
It turns out that this lower bound is actually optimal at the leading and subleading orders in the strong coupling expansion. More surprisingly,
the two-sided numerical bootstrap bounds on the plaquettes averages of the 3D Abelian gauge theories quickly converge to 2D solutions with increasing gauge coupling $\lambda$. We explain below how the Griffiths' inequalities lead to these nontrivial constraints, parallel to the bootstrap method we will study later.

\vspace{3mm}

{\bf The infinite volume limit.} 
 The infinite volume limit of the correlation functions of Abelian lattice gauge theories directly follows from the convex cone $\mQ_{U(1)}$. Consider a finite-size lattice $\Lambda\subset \mathbb{Z}^D$, which is described by an action 
 \begin{equation}
 S_\Lambda=-\frac{2}{\lambda}\sum\limits_{P\in\Lambda}\cos{\theta_P}. 
 \end{equation}
The partition functions and the Wilson loop averages are denoted by $Z_\Lambda$ and $\langle \dots\rangle_\Lambda$. Using two lattices $\Lambda\subset \Lambda'$, assume that their actions differ by certain plaquettes $P\in\bar{\Lambda}\equiv\Lambda'-\Lambda$ with a coefficient $\beta$: $S_{\Lambda'}-S_\Lambda=-\beta\sum\limits_{P\in\Bar{\Lambda}}\cos{\theta_P}\in\mQ_{U(1)}$. With $\beta=0$ we go back to the theory on lattice $\Lambda$.  Now using the Griffiths' second inequality, it gives
 \begin{equation}
    \frac{d}{d\beta}\langle \cos{\theta_\cC} \rangle_{\Lambda'}=\frac{1}{Z^2_{\Lambda'}}\sum\limits_{P\in\Bar{\Lambda}}\langle\cos{\theta_\cC}\cos{\theta_{P}} \rangle_{\Lambda'}-\langle \cos{\theta_\cC} \rangle_{\Lambda'} \langle \cos{\theta_{P}}\rangle_{\Lambda'}\geqslant0. \label{griffiths1}
 \end{equation}
The Wilson loop averages are monotonically increasing with the coupling $\beta\propto 1/\lambda$.
Since the theory on $\Lambda$ is identical to the theory on $\Lambda'$ with $\beta=0$, we have
\begin{equation}
    \langle \cos{\theta}_\cC\rangle_{\Lambda'}\geqslant \langle \cos{\theta}_\cC\rangle_{\Lambda',\;\beta=0}=
    \langle \cos{\theta}_\cC\rangle_{\Lambda}. \label{monot}
\end{equation}
It is know that the Wilson loop averages are also bounded $\langle \cW_\cC\rangle\leqslant 1$, therefore their  limits with infinite lattice $\Lambda\rightarrow \mathbb{Z}^D$ are guaranteed by the monotonicity (\ref{monot}).  

\vspace{3mm}

{\bf A bound on Wilson loop averages in general dimensions.}
Let us consider a theory on a $D$ dimensional lattice $\mathbb{Z}^D$. One can fix the couplings of the plaquettes on a ${D-1}$ dimensional sublattice $\beta_{D-1}=2/\lambda$ while tune the couplings $\beta=0$ for all the other plaquettes. Due to the monotonicity (\ref{monot}), the Wilson loop averages satisfy\footnote{Here we implicitly assume the Wilson loop appears in the sublattice with lower dimensions.} 
\begin{equation}
    \langle \cW_\cC\rangle_{D}\geqslant  \langle \cW_\cC\rangle_{D-1}\geqslant\cdots\geqslant \langle \cW_\cC\rangle_{2}, \label{3D2D}
\end{equation}
i.e., the 2D Abelian lattice gauge theory provides a lower bound for the Wilson loop averages in general dimensions. We will show that the upper bound also converges to the 2D theory with strong gauge coupling $\lambda$.

\section{Bootstrapping 2D $U(1)$ lattice gauge theory} \label{sec4}
Lattice gauge theories and the Ising model on square lattice have been studied using the bootstrap approach \cite{Anderson:2016rcw, Kazakov:2022xuh,Cho:2022lcj}. Strong two-sided bounds have been obtained, while the precision in the physically interested region needs to be improved further. In bootstrap studies the positivity constraints need to be truncated due to the restrictions of computation resource. A fundamental question for the bootstrap method is that with more positivity constraints, can the two-sided bounds keep on converging to the solutions with high precision? In this section, we take the 2D $U(1)$ lattice gauge theory as an example to address this question.
The loop equation (\ref{Z2loopeq}) of the $\mathbb{Z}_2$ lattice gauge theory can be directly solved so no need to use bootstrap.

\subsection{Exact solution of the 2D $U(1)$ lattice gauge theory and Griffiths' inequalities} \label{sec4p2}
The $U(1)$ lattice gauge theory can be solved analytically in 2D \cite{QED2solution, Kogut:1979wt}. Through gauge fixing the 2D loop space is simplified drastically, and one can easily bootstrap the Wilson loops with maximum length up to $L_{\textrm{max}}=60$, therefore it provides an exceptional example to demonstrate how the two-sided bootstrap bounds evolve with increasing $L_{\textrm{max}}$. 

On the 2D square lattice, it is helpful to take the temporal gauge, in which the link variables in the time direction $\mu=\tau$ vanish $\theta_{\tau}(x)=0$.
In this gauge, the link variables $\theta_\ell$ can be replaced by the gauge invariant plaquette variables $\theta_P$ with unit Jacobian \cite{Kogut:1979wt}.
The phase factor in the Wilson loop $\cW_\cC=e^{i\theta_\cC}$ can be expanded through the lattice version of the Stokes law:
\begin{equation}
    \theta_\cC=\sum_{\ell\in\cC}\theta_\ell=\sum\limits_{P\in P_\cC}m_P\theta_P,
\end{equation}
where $P_\cC$ denotes the plaquette enclosed by the Wilson loop path $\cC$, and $m_P$ is the multiplicity of the plaquettes $P$. Negative $m_P$ represents opposite orientation. Then the Wilson loop averages become integral over all independent plaquettes variable.
\begin{equation}
    \left\langle e^{i\theta_\cC} \right\rangle_{\theta_\ell}=\left\langle \prod\limits_{P\in P_\cC}e^{i\theta_P} \right\rangle_{\theta_P}= \prod\limits_{P\in P_\cC}\left\langle e^{i m_P\theta_P} \right\rangle_{\theta_P}, \label{wlthetap}
\end{equation}
in which the subscript in $\langle\dots\rangle_{\theta_X}$ indicates the integral variable is $\theta_X$. The Wilson loop averages are drastically simplified with the plaquettes variables $\theta_P$ as they can be integrated independently. The integrals over $\theta_P$ are labeled by the multiplicity number $m$
\begin{equation}
    I_m(2/\lambda)=\int_{0}^{2\pi}d\theta_P \; e^{\frac{2}{\lambda}\cos{\theta_P} }\cos{m\theta_P},
\end{equation}
where $I_m$ are the incomplete Bessel functions. The Wilson loop averages can be solved in terms of the functions $I_m$, i.e., the plaquette averages is
\begin{equation}
    \langle \cW_P\rangle=\frac{I_1}{I_0}\equiv w_1. \label{plaquette2D}
\end{equation}
For a Wilson loop $\cW_{\cC}$ with area $A$ and the multiplicity of each enclosed plaquettes is $1$, its average is 
\begin{equation}
\langle\cW_{\cC}\rangle=(I_1/I_0)^A\propto \exp{\left(-|\ln{w_1}|A\right)},   
\end{equation}
clear evidence of the confined phase.

Since the functions $I_m(2\lambda)>0$ for $\lambda>0$, the Wilson loop averages are non-negative, consistent with the first Griffiths' inequality. Here we focus on the Griffiths' second inequality, which is actually violated by the solutions (\ref{wlthetap}) of the 2D $U(1)$ lattice gauge theory. To show this, let us
consider two Wilson loops $\cC_1=P^{m_1},\cC_2=P^{m_2}$, both of which are the plaquettes with multiplicities $m_1>0$ and $m_2>0$, respectively. Then we have
\begin{align}
    \langle\cW_{P^{m_1}}\cW_{P^{m_2}}\rangle =\frac{I_{m_1+m_2}}{I_0},~~~~
    \langle\cW_{P^{m_1}}\rangle\langle\cW_{P^{m_2}}\rangle=\frac{I_{m_1}}{I_0}\frac{I_{m_2}}{I_0}.
\end{align}
In the large $\lambda$ limit, the incomplete Bessel functions are
\begin{equation}
    I_m(2/\lambda)=\frac{\lambda^{-m}}{m!}+\frac{\lambda^{-m-2}}{(m+1)!}+\cdots.
\end{equation}
At the leading order, 
\begin{equation}
    I_{m_1}I_{m_2}\simeq \frac{\lambda^{-m_1-m_2}}{m_1!m_2!}> \frac{\lambda^{-m_1-m_2}}{(m_1+m_2)!}\simeq I_0 I_{m_1+m_2},
\end{equation}
which suggests
\begin{equation}
    \langle\cW_{P^{m_1}}\cW_{P^{m_2}}\rangle <\langle\cW_{P^{m_1}}\rangle\langle\cW_{P^{m_2}}\rangle
\end{equation}
for sufficiently large $\lambda$. Numerically, one can check that the above inequality is true even for small $\lambda$s. 
While if the two Wilson loops are of opposite orientations, e.g., $m_2\rightarrow -m_2$, we have
\begin{equation}
    I_{m_1}I_{-m_2}< I_0 ~I_{m_1-m_2}
\end{equation}
and the Griffiths' second inequality is satisfied in this case
\begin{equation}
     \langle\cW_{P^{m_1}}\cW_{P^{-m_2}}\rangle >\langle\cW_{P^{m_1}}\rangle\langle\cW_{P^{-m_2}}\rangle.
\end{equation}
To summarize, the direction of the inequality is affected by the orientation of the Wilson loops, and the Griffiths' second inequality is not always satisfied. This provides an explicit example for the general studies in Section \ref{sec3p2}.

\subsection{Bootstrap bounds on the 2D $U(1)$ lattice gauge theory}
We present a bootstrap study of the 2D $U(1)$ lattice gauge theory. It provides an  example to illustrate the properties of the bootstrap method with a high maximum length $L_{\textrm{max}}$, which are hard to implement for lattice gauge theories in higher dimensions.

As shown in the last section,  after taking gauge fixing for the 2D lattice gauge theory, the Wilson loops factorized into fundamental ingredients $I_m/I_0$ corresponding to the plaquettes with multiplicity $m$. For any Wilson loop along certain path $\cC$, denoting the enclosed plaquettes and their multiplicities by $\{P_i, m_i\}$, its expectation value is
\begin{equation}
    \langle \cW_\cC\rangle=\prod\limits_{i} \left(\frac{I_{m_i}}{I_0}\right)^{|m_i|}. \label{factorize}
\end{equation}
So to solve the theory, it suffices to determine the Wilson loop averages 
\begin{equation}
    w_m\equiv \langle\cW_{P^m}\rangle=I_m/I_0.
\end{equation}
Now we show they are almost uniquely fixed by loop equations and positivity constraints.

\vspace{3mm}

{\bf Loop equations.}
The first step is to solve all the Wilson loop averages $w_{m>1}$ in terms of $w_1$ using loop equations.
The fundamental Wilson loop averages $w_m$ are restricted by the loop equations (\ref{lpequation}), which reduces to a simple form using gauge fixing
\begin{equation}
    w_m-w_{m-2}+(m-1)\lambda w_{m-1}=0,~~w_0=1. \label{lpeq2d}
\end{equation}
Therefore the higher loops $w_{m>1}$ can be fixed recursively, with only one undetermined variable $w_1$, which we fix numerically  using positivity constraints for any given $\lambda$.

\vspace{3mm}

{\bf Positivity constraints.}
There are two types of positivity constraints, the square positivity \cite{Anderson:2016rcw} and reflection positivity \cite{Kazakov:2022xuh}. The later one relates to the disconnected Wilson loops, which due to the factorization of the Wilson loop averages, are expected to be insignificant. So we focus on the square positivity \cite{Anderson:2016rcw}. 

Consider the Hilbert space generated by the Wilson loops $\cW_{P^m}, ~m=0,1,\dots,n$.
The inner product of a general element
\begin{equation}
    \cW=\sum_{i=1}^n c_n \cW_{P^m}
\end{equation}
leads to a semi-definite positive matrix (\ref{sqmatrix}):
\begin{align}
    \cM_{2D}=\left( \begin{array}{ccccc}
       w_0  & w_1 & w_2 & \cdots & w_{n} \\
       w_1  & w_0 & w_1 & \cdots & w_{n-1} \\
       w_2  & w_1 & w_0 & \cdots & w_{n-2} \\
       \vdots  & \vdots & \vdots & \ddots & \vdots \\
       w_n  & w_{n-1} & w_{n-2} & \cdots & w_{0} \\
    \end{array}   \right), \label{sqmatrix2d}
\end{align}
The elements $w_{m>1}$ are replaced by their solutions from the loop equations (\ref{lpeq2d}), which are linear functions of the variable $w_1$. The semi-definite positive condition $\cM_{2D}\succeq0$ provides surprisingly strong constraints on the possible solutions of $w_1$.

\vspace{3mm}

\begin{figure}
\begin{center} 
\includegraphics[width=0.75\linewidth]{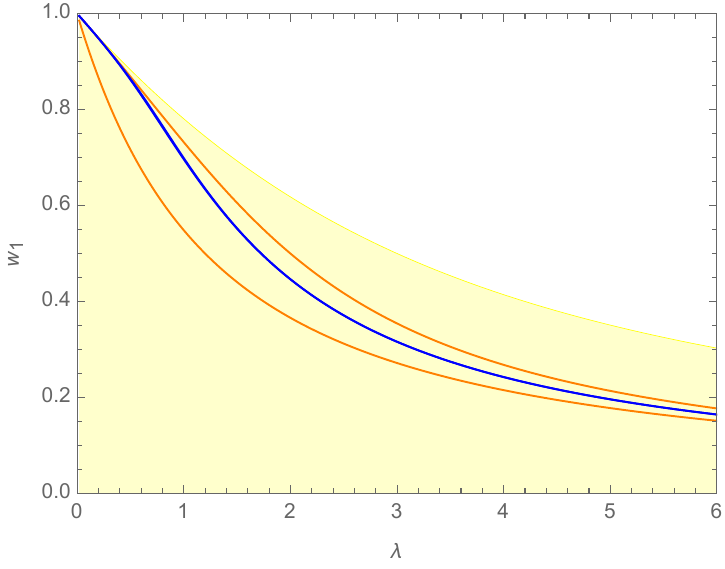}
\end{center} 
\caption{Bounds on the expectation value of the plaquette $w_1$ in 2D $U(1)$ lattice gauge theory with different $L_{\textrm{max}}$. Yellow region: bootstrap allowed region with $L_{\textrm{max}}=8$; orange lines: upper and lower bounds with $L_{\textrm{max}}=12$; blue line(s): upper and lower bounds at $L_{\textrm{max}}=24$. Note at $L_{\textrm{max}}=8$ there is no lower bound above $w_1=0$, while at $L_{\textrm{max}}=24$, the upper and lower bounds are extremely close with each other and indistinguishable in the plot. The bounds at $L_{\textrm{max}}=24$ also coincide with the exact solution (\ref{plaquette2D}) of the 2D $U(1)$ lattice gauge theory.} \label{2dU1plot}
\end{figure}

{\bf Bootstrap bounds. }
It is straightforward to apply the semi-definite optimization program to the matrix $\cM_{2D}$ in (\ref{sqmatrix2d}). The bootstrap bounds on the fundamental variable $w_1$ with different maximum length of the Wilson loops $L_{\textrm{max}}$ are presented in Fig. \ref{2dU1plot}. Here we summarize the interesting properties of the results.

\begin{table} 
\centering 
\caption{The first 10 digits of the bootstrap bounds on the averaged plaquette value $w_1$ in 2D $U(1)$ lattice gauge theory with $L_{\textrm{max}}=60$. }\label{table:L60}
		\begin{tabular}{cccc}
\hline\hline\\[-1em]
	~~	$\lambda$ ~~~~  	&  ~~~~~Lower bound~~~~~ & ~~~~~Upper bound~~~~~~& ~~~~~Exact value~~~~~~ 
 \\[.1em]\hline\\[-1em]
 $0.01$~~ & $0.99749685{\color{red}14} $  &     $0.997496859{\color{red}5} $     &      $0.9974968592 $   
 \\[.1em]\hline\\[-1em]
 $0.1$~~ & $0.97467050{\color{red}40}$  &     $0.974670517{\color{red}2}$     &      $0.9746705078$   
 \\[.1em]\hline\\[-1em]
			$0.2$~~ & $0.9485998{\color{red}188}$  &     $0.9485998{\color{red}342}$     &      $0.9485998259$   
 \\[.1em]\hline\\[-1em]
		$0.4$~~	  & $0.89338313{\color{red}27}$   &   $0.8933831{\color{red}412}$    & $0.8933831370$    \\[.1em]\hline\\[-1em]
		$0.6$~~	  & $0.83190007{\color{red}05}$   &   $0.831900071{\color{red}7}$    & $0.8319000711$  \\[.1em]\hline\\[-1em]
		$0.8$~~	  & $0.764996747{\color{red}0}$   &   $0.76499674{\color{red}81}$    & $0.7649967475$  \\[.1em]\hline\\[-1em]
		$1.0$~~	  & $0.697774657{\color{red}5}$   &   $0.69777465{\color{red}83}$    & $0.6977746579$  \\[.1em]\hline\\[-1em]
		$2.0$~~	  & $0.446389965{\color{red}6}$   &   $0.44638996{\color{red}61}$    & $0.4463899659$  \\[.1em]\hline\\[-1em]
		$4.0$~~	  & $0.2424996125$   &   $0.242499612{\color{red}6}$   & $0.2424996125$  \\[.1em]\hline\\[-1em]
		$6.0$~~	  & $0.1643939155$   &  $0.1643939155$    & $0.1643939155$                 
 \\[.1em]\hline\hline 
		\end{tabular}
\end{table}

\begin{figure}
\includegraphics[width=0.49\linewidth]{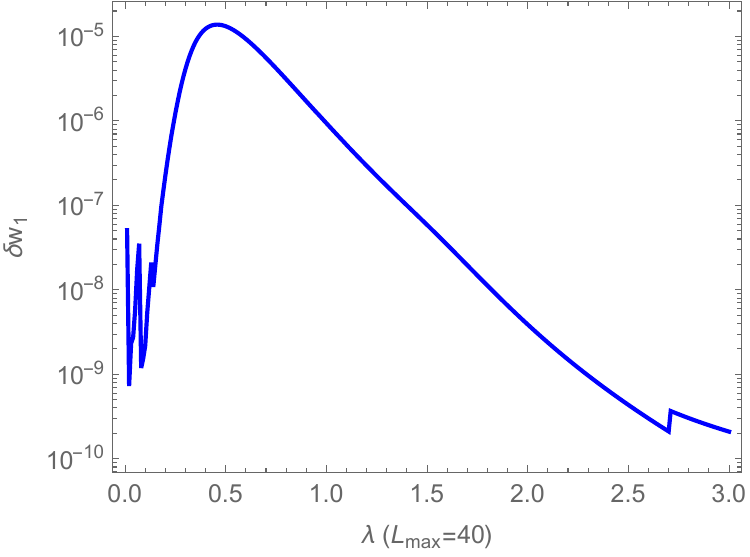}~
\includegraphics[width=0.49\linewidth]{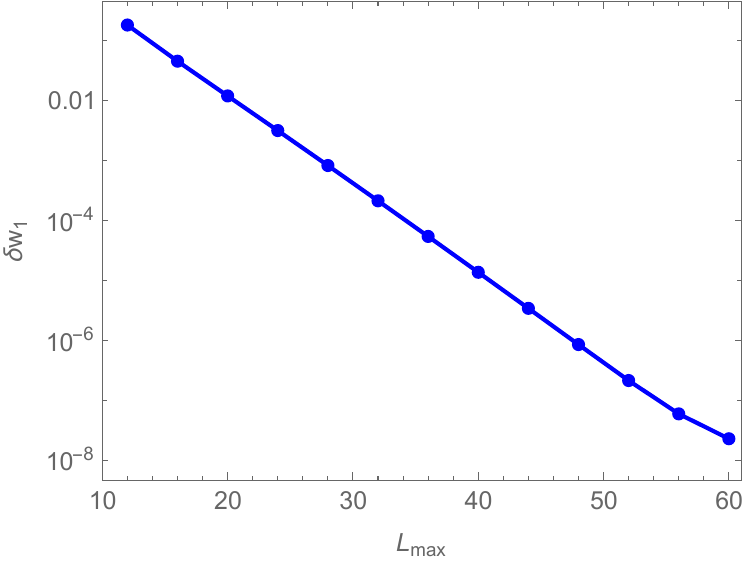}
\caption{Left panel: precision ($\delta w_1$) of the two-sided bounds on $w_1$ at $L_{\textrm{max}}=40$ for different $\lambda$; right panel: maximum value of $\delta w_1$ for each $3\leqslant L_{\textrm{max}}/4\leqslant15$.} \label{BDL40}
\end{figure}

The bounds with low $L_{\textrm{max}}$ are reminiscent to the bootstrap bounds on the $SU(\infty)$ lattice gauge theories in \cite{Kazakov:2022xuh}. Specifically, There is a weak upper bound with $L_{\textrm{max}}=8$, while there is no nontrivial lower bound in this case. 
The bounds are improved notably at $L_{\textrm{max}}=12$. But still, like the the results in \cite{Anderson:2016rcw, Kazakov:2022xuh}, not strong enough to extract precise dynamical information. In previous bootstrap studies \cite{Anderson:2016rcw,Kazakov:2022xuh,Cho:2022lcj},  it is expected that by introducing more positivity constraints, the bootstrap bounds should be significantly improved. While in previous bootstrap studies, it requires much more computation resource to extend the bootstrap implements to larger systems so it is hard to verify this conjecture. Now for 2D $U(1)$ lattice gauge theory, the bootstrap implementations are remarkably simpler thanks to the gauge fixing. The bootstrap implementation can be straightforwardly extended to higher $L_{\textrm{max}}$. In Fig. \ref{2dU1plot} the upper and lower bounds on $w_1$ with $L_{\textrm{max}}=24$ are shown, which are too close with each other and become indistinguishable. They also nearly coincide with the analytical solution (\ref{plaquette2D}) within the error $3\times 10^{-3}$. 

We have computed the bootstrap bounds up to $L_{\textrm{max}}=60$ and studied  the precision of bootstrap bounds with increasing $L_{\textrm{max}}$. The first ten digits of the upper and lower bounds with $L_\textrm{max}=60$ are shown in Table. \ref{table:L60}, in which the numerical precision is of order $10^{-8}$ for $\lambda\sim0.2$ and becomes even better for other $
\lambda$s. 
The numerical precision for different $L_\textrm{max}$ is shown in Fig. \ref{BDL40}. In the left panel, we show the precision of the two-sided bounds for different $\lambda$. The two-sided bounds converge much better in the strong and weakly coupled regions. The precision becomes weaker for the intermediate $\lambda$. In the right panel of Fig. \ref{BDL40}, we show the maximum spread, or the precision of the two-sided bounds for fixed $L_{\textrm{max}}$. The precision 
$\delta w_1$ improves linearly with increasing $L_{\textrm{max}}$. At $L_{\textrm{max}}=60$ it reaches a remarkably high precision $10^{-8}$!\,\footnote{The bounds in Figs. \ref{2dU1plot} and \ref{BDL40}  are computed using the ``{\it SemidefiniteOptimization}" program in {\it Mathematica},  with ``Tolerance" $10^{-8}$. The precision at $L_{\textrm{max}}=60$ is close to the ``Tolerance" value so the results are partially affected by the numerical setup.} It clearly shows that the precision of the two-sided bootstrap bounds is proportional to the maximum length of the Wilson loops $L_{\textrm{max}}$: $-\ln(\delta w_1)\propto L_{\textrm{max}}$.

We summarize the gist of this section. The bootstrap bounds on the 2D $U(1)$ lattice gauge theory provide strong evidence indicating that {\it with sufficiently large $L_{\textrm{max}}$, the loop equations and the positivity conditions can generate extremely precise numerical solutions for this lattice gauge theory! }

\section{Bootstrapping 3D Abelian lattice gauge theories} \label{sec5}
In this section we study the Abelian gauge theories on a 3D lattice. The Abelian lattice gauge theories can have a gauge group $U(1)$ or its subgroups $\mathbb{Z}_N$. It is known that the 3D $\mathbb{Z}_2$ lattice gauge theory is dual to the 3D Ising model \cite{Wegner:1971app}. The theory is confined with strong coupling while deconfined in the weakly coupled region, and there is a phase transition between the confined and deconfined phases. On the other hand, the 3D $U(1)$ lattice gauge theory is confined for all gauge coupling \cite{Polyakov:1975rs,Polyakov:1976fu,Gopfert:1981er}. An interesting question is to understand how the $\mathbb{Z}_N$ lattice gauge theories approach the $U(1)$ lattice gauge theory in the large $N$ limit, see \cite{Luo:2023cjv} for a recent Monte Carlo study. It would be tempting to know if bootstrap method can shed new light for these well known theories. We start with the two most classical examples: the $U(1)$ and $\mathbb{Z}_2$ lattice gauge theories.  

We follow the previous bootstrap implementations developed in \cite{Anderson:2016rcw,Kazakov:2022xuh}. 
In particular, we prepare the set of Wilson lines and take their symmetry reductions according to the methods in \cite{Kazakov:2022xuh}.
In Section \ref{sec4p2}, we have shown that in 2D the bootstrap implementation can be drastically simplified by gauge fixing, through which one can directly compute the correlation functions using the plaquette variables. In 3D, the plaquette variables are restricted by the Gauss' law and so cannot be independent integration variables.

\subsection{Bootstrapping the 3D $U(1)$ lattice gauge theories}
The 3D $U(1)$ lattice gauge theory has been shown to be confined in the long distance limit for any gauge coupling due to the condensation of the monopoles \cite{Polyakov:1975rs,Polyakov:1976fu}. It provides a theoretically tractable example for studying the general properties of the confinement mechanism, including the role of topological gauge field excitation, the string tension and glueball spectrum, etc. 
There are abundant perturbative results for the $U(1)$ lattice gauge theories in general dimensions. In particular, the plaquette average $u_P\equiv \langle \cW_P\rangle$ has been computed to high loops in different limits. The weak coupling expansion of $u_P$ has been computed to the order $\lambda^4$ in \cite{HORSLEY1981290}:
\begin{equation}
    u_P=1-\frac{1}{6}\lambda-\frac{1}{72}\lambda^2-0.0041375 \lambda^3-0.000175 \lambda^4+O(\lambda^5).\label{weakup}
\end{equation}
The strong coupling expansion of $u_P$ has been computed to the order $\lambda^{-15}$ in \cite{PhysRevD.11.2104}:
\begin{equation}
    u_P=\frac{1}{\lambda }-\frac{1}{2 \lambda ^3}+\frac{7}{3 \lambda ^5}-\frac{395}{48 \lambda ^7}+\frac{1173}{40 \lambda ^9}-\frac{507803}{4320 \lambda ^{11}}+\frac{7352027}{15120 \lambda ^{13}}-\frac{443004913}{215040 \lambda ^{15}}+O(\lambda^{-17}). \label{strongup}
\end{equation}
Recently, the 3D $U(1)$ lattice gauge theory has been studied using quantum simulation  \cite{Zohar:2012ts, Paulson:2020zjd}.

\begin{figure}
\begin{center} 
\includegraphics[width=0.8\linewidth]{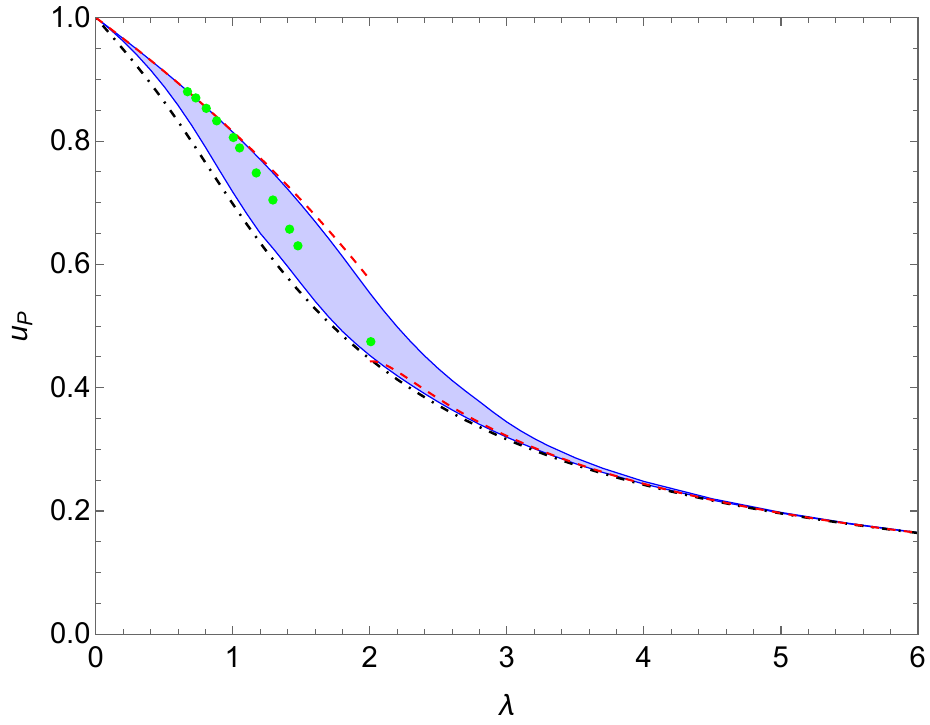}
\end{center} 
\caption{Bootstrap bounds on the plaquette average $u_P$ in the 3D $U(1)$ lattice gauge theory. Blue region: bootstrap allowed region with $L_{\textrm{max}}=16$; black dot-dashed line: analytical solution (\ref{plaquette2D}) of the 2D $U(1)$ lattice gauge theory $w_1$; red dashed lines: strong and weak coupling expansions of the 3D $U(1)$ lattice gauge theory; green dots: Monte Carlo results \cite{Loan:2002ej}. } \label{3DU1bd}
\end{figure}

\begin{figure}
\includegraphics[width=0.49\linewidth]{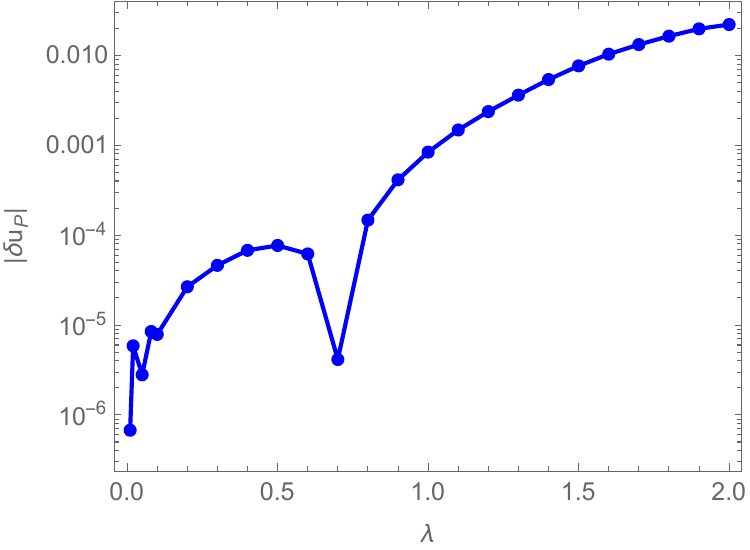}~
\includegraphics[width=0.49\linewidth]{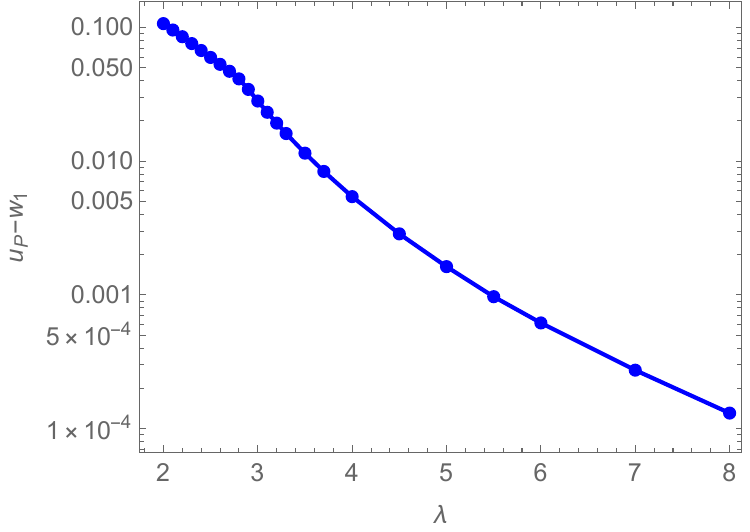}
\caption{Left panel: difference between bootstrap upper bound $u_P$ and weak coupling expansion $u_P^w$ of the plaquette average $\delta u_P=|u_P-u_P^{w}|$; right panel: difference between bootstrap upper bound $u_p$ and 2D analytical solution $w_1$ of the plaquette average $ u_P-w_1$.} \label{3DU1dif}
    \end{figure}

We present the bootstrap bounds on the 3D $U(1)$ lattice gauge theory in Fig. \ref{3DU1bd}. The bootstrap results are computed with maximum Wilson loop length $L_{\textrm{max}}=16$. The blue shadowed region denotes the bootstrap allowed region for the expectation value of the plaquette $u_P$.  The two-sided bounds on $u_P$ are well converged in the strong or weak coupling limit. The bounds monotonically decrease with increasing $\lambda$, as predicted by the Griffiths' inequalities (\ref{griffiths1}). We compare the bootstrap bounds with the perturbative results (red dashed lines in Fig. \ref{3DU1bd}) in the strong or weak coupling region.  In the weak coupling limit, the upper bound on $u_P$ is remarkably close to the weak coupling expansion prediction at the order $\lambda^4$ (\ref{weakup}). The absolute value of their difference is shown in the left panel of Fig. \ref{3DU1dif}. While in the strong coupling limit, both the upper and lower bounds converge to the large coupling expansion of $u_P$ (\ref{strongup}). In the range with intermediate $\lambda$, the bootstrap bound are not well converged yet for $L_{\textrm{max}}=16$. Nevertheless, they are well consistent with the Monte Carlo results \cite{Loan:2002ej} given by the green dots in the figure.  

Another interesting point in Fig. \ref{3DU1bd}
is that the lower bound on $u_P$ is quite close to the 2D analytical solution of the plaquette average $w_1$ (\ref{plaquette2D}), denoted by the black dot-dashed line in the figure.  Especially in the region with large coupling coefficient $\lambda$, the lower bound on $u_P$ almost coincides with the analytical solution of $w_1$ given by the black dash-dotted line, and the upper bound on $u_P$ tends to merge with $w_1$ too! In the right panel of Fig. \ref{3DU1dif}, we show the difference between the upper bound and the 2D analytical solution $u_{P}-w_1$, which decreases exponentially with increasing $\lambda$. The bootstrap bounds are above the average plaquette in 2D, consistent with the lower bound (\ref{3D2D}) on $u_P$ obtained from the Griffiths' inequalities. While the bootstrap result suggests that the lower bound obtained from Griffiths' inequalities is actually almost optimal: the upper bound is quite close to the 2D solution with $\lambda\geqslant6$. 

The universal behavior of the plaquette average in the strong coupling limit in different dimensions is also shown implicitly in the perturbative expansion. In \cite{HORSLEY1981290} the authors have computed the free energy of the $U(1)$ lattice gauge theory on $D$ dimensional lattice, which corresponds to the plaquette average value:
\begin{equation}
    u_P=\frac{1}{\lambda}-\frac{1}{2\lambda^3}+\frac{6D-11}{3\lambda^5}-\frac{384 D-757}{96\lambda^7}+O(\lambda^{-9}).
\end{equation}
The leading and sub-leading order terms are independent of $D$. Similar results also appear in the $SU(2)$ Yang-Mills lattice gauge theory \cite{HORSLEY1981290}, in which dimension-dependent corrections only appear at the order $\lambda^{-5}$ or higher.
Physically, this indicates that in the strongly coupled region, the gauge interactions are confined in a small region, and are not sensitive to the configuration of the whole lattice system. It will be interesting for future studies to understand more universal behaviors and their corrections in confinement.

\subsection{Bootstrapping the 3D $\mathbb{Z}_2$ lattice gauge theories}
We bootstrap another widely interested Abelian lattice gauge theory, the 3D $\mathbb{Z}_2$ lattice gauge theory.  This theory provides the first example for phase transitions without local order parameter \cite{Wegner:1971app} and plays an important role in the recent studies of generalized global symmetries \cite{Shao:2023gho}. We want to explore the role of positivity in determining the dynamical information of this theory. Its dual version, the 3D Ising model on the cubic lattice has been bootstrapped in \cite{Cho:2022lcj}, in which interesting two-sided bounds have been generated. 

\begin{figure}
    \centering
    \includegraphics[width=1\linewidth]{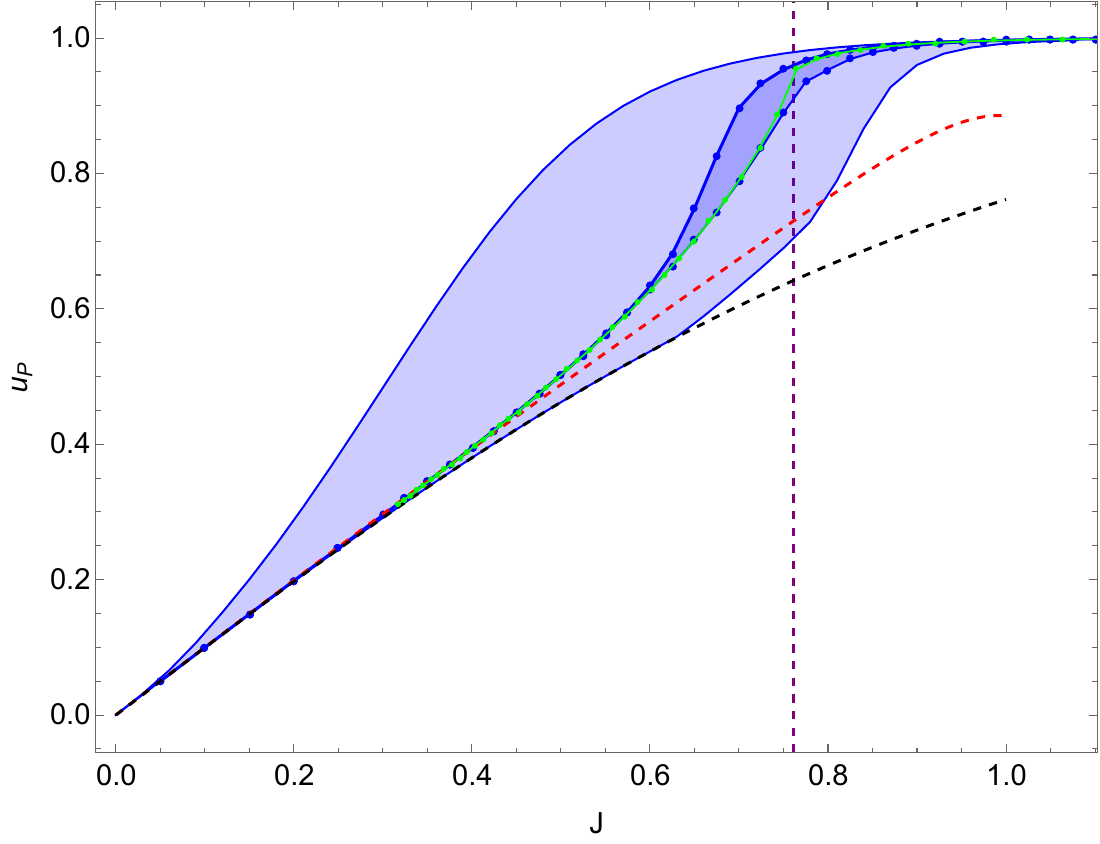}
    \caption{Bootstrap bounds on the plaquette average $u_P$ in the 3D $\mathbb{Z}_2$ lattice gauge theory. The light (dark) blue shadowed region denotes the bootstrap allowed region for $u_P$ with maximum Wilson loop length $L_{\textrm{max}}=12$ ($L_{\textrm{max}}=16$). The green dots are from Mento Carlo data for the 3D Ising model \cite{Cho:2022lcj}, which has been transferred to the 3D $\mathbb{Z}_2$ lattice gauge theory through the duality relation (\ref{Z2Ising}). The red dashed line represents the strong coupling expansion of $u_P$ to the order $J^{15}$. Solution of the 2D $\mathbb{Z}_2$ lattice gauge theory $u_P=\tanh{J}$ is denoted by the black dashed line. The purple vertical dashed line corresponds to $J=J^*\approx 0.761413$, where the confinement-deconfinement phase transition happens.}
    \label{fig:3DZ2bd}
\end{figure}

The bootstrap bounds on the plaquette average $u_P$ in 3D $\mathbb{Z}_2$ lattice gauge theory are shown in Fig. \ref{fig:3DZ2bd}. The two-sided bounds on $u_P$ can be obtained with $L_{\textrm{max}}=12$ (light blue shadowed region), and it quickly shrinks to a much narrower region with $L_{\textrm{max}}=16$ (dark blue shadowed region). Remarkably, the precision of the two-sided bounds is significantly improved for small or large $J\notin(0.6, 0.8)$, and the error bar of the two-sided bounds is $\delta u_P\approx10^{-2}$ or smaller. 
We compare the bootstrap bounds with the strong coupling expansion up to the order $J^{15}$ \cite{PhysRevD.11.2104}:
\begin{equation}
    u_P=J-\frac{J^3}{8}+\frac{7 J^5}{48}-\frac{395 J^7}{3072}+\frac{1173 J^9}{10240}-\frac{507803 J^{11}}{4423680}+\frac{7352027 J^{13}}{61931520}-\frac{443004913 J^{15}}{3523215360}+O(J^{17})
\end{equation}
which is represented by the red dashed line in the figure. In the strong coupling limit (small $J$), the two-sided bootstrap bounds indeed coincide with the perturbative result with high precision.
Another interesting property is that for $J<0.6$, the lower bound is close to the exact 2D solution of the $\mathbb{Z}_2$ lattice gauge theory: $u_P=\tanh{J}$,  given by the black dashed line in the figure. In fact, with $L_{\textrm{max}}=12$, the lower bound on $u_P$ almost coincides with the exact 2D solution for $J<0.62$. This confirms again the robustness of the lower bound on $u_P$ from the Griffiths' inequalities (\ref{3D2D}), and the modern bootstrap algorithm can move much further. 

Different from the 3D $U(1)$ lattice gauge theory, the 3D $\mathbb{Z}_2$ lattice gauge theory admits a second order phase transition corresponding to the deconfinement process. It is well known that the  3D $\mathbb{Z}_2$ lattice gauge theory is dual to the 3D Ising model. Denote the coupling coefficient of the dual 3D Ising model by $J'$, the free energies of the two theories satisfy
\begin{equation}
    F_{\textrm{Ising}}(J')=F_{\mathbb{Z}_2 \textrm{ gauge}}(J)-\frac{3}{2}\ln \sinh(2J)+\frac{1}{2}\ln 2. \label{Z2Ising}
\end{equation}
By taking derivatives with respect to $J/J'$, above duality relation leads to an equation between the plaquette average $u_P$ and the nearest-spin correlator $\langle s_i s_{i+\mu}\rangle$ in the 3D Ising model. We use the Monte Carlo data on the nearest-spin correlator $\langle s_i s_{i+\mu}\rangle$ presented in \cite{Cho:2022lcj} to compute the plaquette average $u_P$. The results are given by the green dots in Fig. \ref{fig:3DZ2bd}, which show excellent agreements with the bootstrap bounds with $L_{\textrm{max}}=16$.
The critical coupling coefficient ${J'}^*$ for the 3D Ising model has been estimated to extremely high precision using large-scale Monte Carlo simulation \cite{Ferrenberg_2018}:
\begin{equation}
    {J'}^*=0.221 654 626 (5 ).
\end{equation}
Applying the duality relation between the coupling coefficients of the two theories
\begin{equation}
\sinh{2J}\,\sinh{2J'}=1,
\end{equation}
one gets the critical coupling coefficient for the 3D $\mathbb{Z}_2$ lattice gauge theory 
\begin{equation}
    J=J^*\approx 0.761413292 (11)  .
\end{equation}
We denote $J^*$ by a purple vertical line in Fig. \ref{fig:3DZ2bd}. Near the phase transition point $J^*$, the precision of the two-sided bound becomes weaker $\delta u_P\sim 0.1$.  Presumably the spread of the two-sided bound relates to the long range correlation near the continuous phase transition,\footnote{The spread of the bootstrap bounds near certain coupling coefficient seems to be related to more reasons besides the long range correlation near the fixed point, since a similar spread appears in Fig. \ref{3DU1bd}, the bootstrap bounds of the $U(1)$ lattice gauge theory in which no phase transition occurs. } for which larger Wilson loops with much higher $L_{\textrm{max}} > 16$ are needed to capture the dynamical information and pin down the theory numerically. 
Motivated by the relation between the precision of the two-sided bound and the maximum length $L_{\textrm{max}}$ shown in Fig. \ref{3DU1dif}, it is promising that with much higher $L_{\textrm{max}}$ the bootstrap constraints could provide sufficiently strong constraints on the theory even at $J^*$.
The next step following this work is to bootstrap the 3D $\mathbb{Z}_2$ lattice gauge theory with $L_{\textrm{max}} =20, 24$, and it would be very interesting to see the new bootstrap bounds near $J^*$.

 \subsection{Towards bootstrapping the string tension and glueball masses}
Previous bootstrap computations focused on the fundamental observable, the expectation value of the plaquette.  It is important to know the averages of the larger Wilson loops to determine the phases in the long distance limit, in particular the string tension and glueball mass for the confined phase. In this section, we present primary results towards bootstrapping the string tension and glueball masses.

We take the 3D $U(1)$ lattice gauge theory as an example. By choosing a suitable limit $1/a e^2\rightarrow \infty$ while fixing the mass gap, the theory is confined in the infrared with free massive bosons. The string tension and glueball mass spectrum have been studied using perturbative and Monte Carlo approaches. In particular the non-perturbative Monte Carlo simulation confirms the results from semi-classical analysis, though the numerical coefficients are notably different.  Overall this theory provides an ideal example for the bootstrap studies of gauge confinement.

\vspace{3mm}

{\bf String tension. }
In the confined phase, the large rectangle Wilson loop $\cW[I,J]$ with two sides of lengths $I$ and $J$ is dominated by an area law
\begin{equation}
    W[I,J]\equiv\langle\cW[I,J]\rangle\sim e^{-\sigma IJ},
\end{equation}
where $\sigma$ is the string tension. While in practice one works on the Wilson loops with finite $I,J$,  and there are self-energy contributions (proportional to the perimeter $I+J$)  to $W[I,J]$:
\begin{equation}
    W[I,J]\simeq e^{-\sigma IJ-b(I+J)-a}.
\end{equation}
The string tension $\sigma$ can be extracted using the Creutz ratios \cite{Creutz:1980zw}
\begin{equation}
    \chi(I,J)=-\ln{\left(\frac{W[I,J]~W[I-1,J-1]}{W[I,J-1]~W[I-1,J]}\right)}. \label{creutz}
\end{equation}
In our bootstrap implementation with $L_{\textrm{max}}=16$, the maximum rectangle Wilson loop is $\cW[3,2]$. We estimate the Creutz ratio $\chi(3,2)$ based on the bootstrap bounds on the Wilson loops appearing in (\ref{creutz}) with $I=3, J=2$.

\begin{table} 
\centering 
\caption{Bootstrap bounds on the Wilson loops in (\ref{creutz}) at different $\lambda$. }\label{table:st}
		\begin{tabular}{ccccccc}
\hline\hline\\[-1em]
		~~$\lambda$~~	&  ~~$W[3,2]$~~ & ~~$W[3,1]$~~& ~~$W[2,2]$~~ & ~~$W[2,1]$~~ & ~~~$\chi_{\textrm{ext}}$~~~ & ~~MC\cite{Loan:2002ej}~~ 
 \\[.1em]\hline\\[-1em]
			$0.6$  &     0.57(16)     &      0.71(8)    &  
      0.66(10) & 0.79(6) & 0.04 &   N.A.
 \\[.1em]\hline\\[-1em]
		$0.8$	  &  0.44(26)  &  0.60(14)     & 0.53(17) & 0.70(11) & 0.05 & 0.010 
 \\[.1em]\hline\\[-1em]
		$1.0$	  &  0.34(34)  & 0.49(20)& 0.42(23) & 0.61(16)                &      0.06     &    0.050     
 \\[.1em]\hline\hline 
		\end{tabular}
\end{table}

In Table \ref{table:st} we show the bootstrap bounds on the Wilson loop averages. Though the bounds on $W[I,J]$ are two-sided and strict, the precision with  given $\lambda$s are not high enough to generated a useful strict bounds on $\chi$.  We present estimations on $\chi$ (denoted by $\chi_{\textrm{ext}}$) based on the extremal solutions of the Wilson loop averages near the upper bounds on $u_P$. At $\lambda=1.0$ the estimated $\chi$ is close to the Monte Carlo result \cite{Loan:2002ej}, though there are notable errors in other estimations. Above primary analysis indicates that with stronger bootstrap bounds on the Wilson loop averages, one should be able to provide data on the string tension with impressive precision.  

\vspace{3mm}

{\bf Glueball mass.}
The glueball mass is a fascinating problem in the studies of gauge confinement and is much more difficult to evaluate. The method to evaluate the glueball spectrum for the lattice gauge theories has been developed for decades, see e.g. \cite{Teper:1998te}.
The idea is to construct a state $|\cO(\tau)\rangle$, a superposition of Wilson loops on a fixed time slice, which has zero momentum and proper quantum numbers. 
The state $|\cO\rangle$ is carefully prepared so that it has large overlap with the glueball states.
Then the glueball spectrum can be extracted from the exponential decay of the correlation function $\langle \cO(\tau)\cO(0)\rangle$. The correlator can be expanded in terms of the expectation values of the disconnected Wilson loops $\langle \cW_1(\tau)\cW_2(0)\rangle$. In principle, one can bootstrap above correlators and generate strict two-sided bounds on the glueball spectrum too. A critical question is the precision of the Wilson loop averages. 

\begin{figure}
	 
		\includegraphics[width=0.15\linewidth]{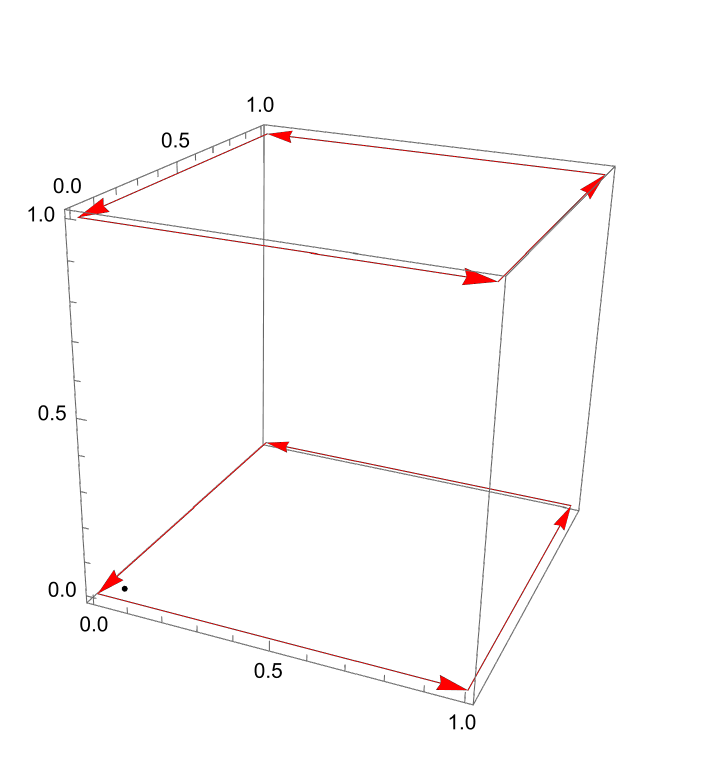}
		~~~~~~\includegraphics[width=0.13\linewidth]{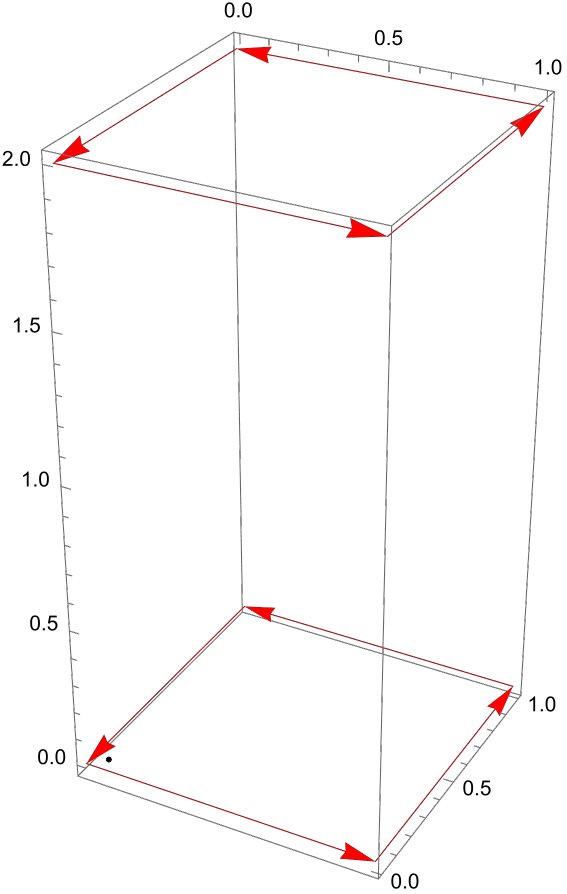}
~~~~~\includegraphics[width=0.26\linewidth]{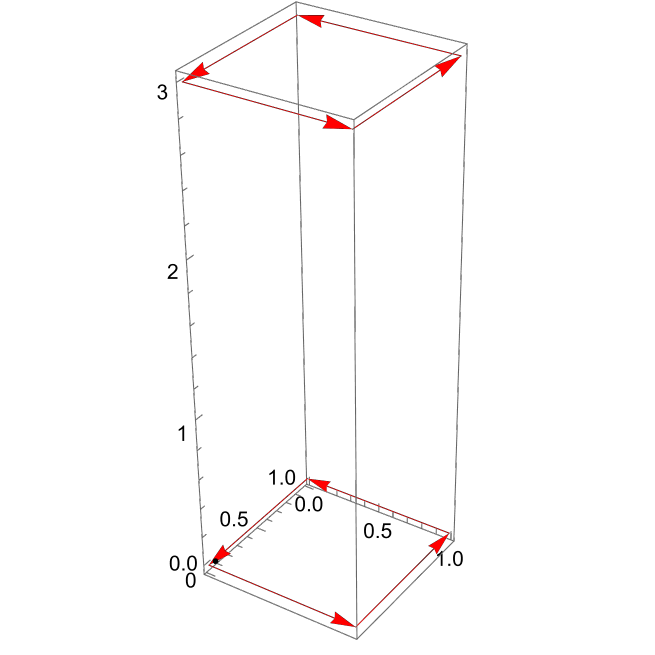}
\includegraphics[width=0.32\linewidth]{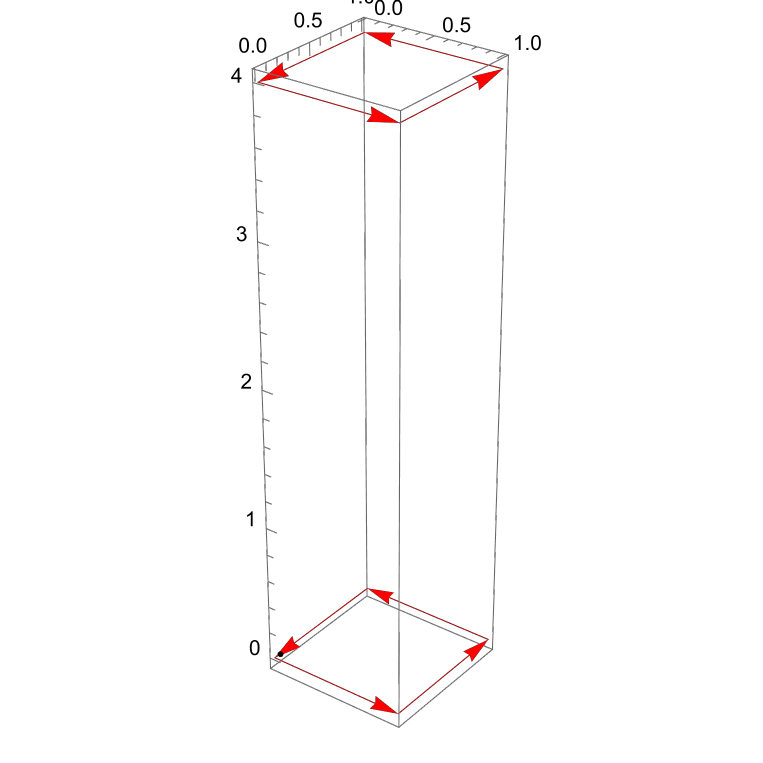}
	 
	\caption{ Correlators of the plaquettes with different temporal (vertical direction) separations.
	} \label{fig:discWLs}
\end{figure}
\begin{table} 
\centering 
\caption{Bootstrap bounds on the correlators of two (anti)parallel plaquettes $\langle P(\tau)P(0) \rangle$ $\left(\langle P(\tau)^*P(0) \rangle\right)$ at $\lambda=1$. }\label{table:gb}
		\begin{tabular}{ccccc}
\hline\hline\\[-1em]
		 	&  ~~~$\tau=1$~~~ & ~~~$\tau=2$~~~& ~~~$\tau=3$~~ & ~~~$\tau=4$~~~~  
 \\[.1em]\hline\\[-1em]
			$\langle P(\tau)P(0)\rangle$~~ & 0.56(12)  &     0.58(14)     &      0.59(15)   & 0.61(22) 
 \\[.1em]\hline\\[-1em]
		$\langle {P}(\tau)^* P(0)\rangle$~~	  & 0.64(15)   &   0.60(15)    & 0.59(15) &  0.63(22)                   
 \\[.1em]\hline\hline 
		\end{tabular}
\end{table}

We consider the simplest correlators of the  Wilson loops on two different time slices, as shown in Fig. \ref{fig:discWLs}.
In the figure, the correlators consist of two parallel plaquettes with the same orientation: $\langle P(\tau)P(0)\rangle$. The strict two-sided bounds on the averages of these correlators are presented in Table \ref{table:gb}. For comparison we also show the estimates of the correlators of the two anti-parallel plaquettes. Similar two-sided bounds can be obtained for correlators of more general Wilson loops on two different time slices. 
The glueball mass spectrum is hidden in these two-sided bounds, which could be carved out given the precision of the two-sided bounds can be improved significantly.

\section{Conclusion and Discussion} \label{sec6}
In this work, we studied the Abelian lattice gauge theories using the bootstrap method, by exploiting the loop equations of the Wilson loop averages associated with the positivity conditions. Our bootstrap results can be summarized as follows:
\begin{itemize}
    \item We addressed a fundamental question on bootstrapping the lattice gauge theories:
    \newline {\it Can the loop equations and positivity conditions be sufficient to numerically pin down a lattice gauge theory?} \newline We studied this question based on the 2D $U(1)$ lattice gauge theory, in which the loop equations are drastically simplified after taking gauge fixing. We explored the bootstrap constraints for Wilson loops up to length $L_{\textrm{max}}=60$, which generates two-sided bootstrap bounds with remarkably high precision $\delta w_1\simeq 10^{-8}$. We observed that the numerical precision of the two-sided bounds  increases linearly with higher $L_{\textrm{max}}$: $\ln{(\delta w_1)}\propto L_{\textrm{max}}$. The results lead to a persuasive conclusion: \newline {\it With large $L_{\textrm{max}}$, the loop equations and positivity conditions can numerically pin down the 2D $U(1)$ lattice gauge theory.}
    \item We bootstrapped two classical Abelian lattice gauge theories: the $U(1)$ and $\mathbb{Z}_2$ lattice gauge theories on the 3D cubic lattice. We obtained two-sided bounds on the plaquette average. The two-sided bounds quickly converge in the strong or weak coupling limit with high precision. We compared our numerical results with the perturbative expansions in the strong or weak coupling limit, as well as the Monte Carlo data and found excellent agreements. The numerical precision of the two-sided bounds is less impressive for certain intermediate coupling coefficients, which we expect can be significantly improved by bootstrapping Wilson loops with higher $L_{\textrm{max}}$.
    \item We found interesting connections between the bounds from bootstrap constraints and Griffiths' inequalities. The real sector of the Abelian lattice gauge theories forms a multiplicative convex cone and satisfies the Griffiths' inequalities, which further lead to a lower bound on the expectation values of the Wilson loops. The two-sided bootstrap bounds converge to the Griffiths lower bound with large coupling coefficient, suggesting that the Griffiths lower bound is actually optimal in strong coupling limit.
    \item We explored the bootstrap constraints on the string tension and glueball mass for the 3D $U(1)$ lattice gauge theory. We obtained two-sided bounds on the expectation values of larger Wilson loops and the correlators of plaquettes on different time slices, which are needed to estimate the string tension and glueball mass spectrum in the confined phase. The precision of the two-sided bounds needs to be  improved significantly to extract the key information for the fascinating problem--gauge confinement.  
\end{itemize}

\vspace{1mm}
We discuss problems and future studies inspired by the results in this work. 

\vspace{1mm}
Perhaps the most important problem is to extend the 2D lattice bootstrap with high $L_{\textrm{max}}$ to higher dimensions. 
The 2D $U(1)$ results  are  encouraging for bootstrapping higher dimensional lattice gauge theories. It is tempting to believe that for higher dimensional lattice gauge theories, with sufficiently large $L_{\textrm{max}}$ the bootstrap bounds may also converge to solutions with high precision. Convergence of the bootstrap bounds on the lattice Ising model has been studied in \cite{Cho:2023ulr}.  While this is hard to be verified in practise, as the bootstrap implementations quickly become too complicated to study. The number of Wilson loops increases exponentially with higher $L_{\textrm{max}}$, but empirically only small part of them have notable contributions on the bootstrap bounds.
It would be important to find systematical selection rules for the  Wilson loops to improve the numerical efficiency. This problem gets more subtle as the Wilson loops are entangled with each other through loop equations. 

Another interesting extension of our 2D bootstrap results is to consider the 2D gauge theories coupled with matter fields. Recently these theories have been studied using different approaches, see e.g.  \cite{Dempsey:2023gib,Dempsey:2023fvm,Aharony:2023tam,  Cherman:2024onj,Bergner:2024ttq}. From the lattice bootstrap point of view, the gauge fixing can help to simplify the loop equations, with new ingredients from the matter fields. It would be interesting to see if the 2D lattice gauge theories coupled to matters can still be effectively bootstrapped with high $L_{\textrm{max}}$. We leave this problem for  future study.

It is interesting that with a large coupling coefficient $\lambda$, the 3D two-sided bootstrap bounds merge and coincide with the 2D theories. It has been predicted using the Griffiths' inequalities long ago that the 2D theories provide lower bounds for the Wilson loop averages in higher dimensions \cite{DeAngelis:1977sr}. While our bootstrap bounds
suggest the Griffiths lower bound is actually optimal with large couplings. Physically this may relate to the fact that with strong coupling the gauge interactions are confined in a small region so are insensitive to the global geometry of the lattice. Nevertheless, it is remarkable that the result can be obtained just from positivity/convexity reasons. It would be interesting to explore the constraints from the positivity and convexity further and study the universal behaviors of the confinement. Another interesting question is whether there are similar properties for the Yang-Mills gauge theories. The persumed convex cones in the Yang-Mills lattice gauge theories are quite subtle. One may firstly gain evidence on this problem by comparing the bootstrap bounds of the same gauge group but in different dimensions \cite{Anderson:2016rcw,Kazakov:2022xuh, Kazakov:2024ool}, and check if the higher dimensional bounds are strictly above while converging to the 2D solutions with large coupling.   

We presented the primary bootstrap bounds which can be used to estimate the string tension and glueball mass spectrum. Interesting data on these parameters could be obtained given the precision of the bootstrap bounds can be improved by one order better. Motivated by the 2D bootstrap results on the precision $\delta w_1$ and the maximum length of the Wilson loops $L_{\textrm{max}}$, we expect the precision of the 3D bootstrap bounds could be significantly improved by increasing $L_{\textrm{max}}$ from our current value $16$ to $24$. We leave these explorations for future work.

\section*{Acknowledgements}
We would like to thank Zechuan Zheng for very helpful discussions and communications. We are grateful to Jiaxin Qiao for the comments on the first version of this paper.
This research was supported by the Startup Funding  4007022314 of the Southeast University, and the National Natural Science Foundation of China funding No. 12375061.

\bibliographystyle{utphys.bst}
\bibliography{AbelianLGT}
\end{document}